%% file: StatCorrRandomMeasurements.tex
\documentclass[10pt,twocolumn,pra]{revtex4-1}
\usepackage{graphicx}
\usepackage{amsthm,amsfonts,amstext,amssymb,amsmath}
\usepackage{latexsym}
\usepackage[margin,index]{fixme}
\usepackage[utf8]{inputenc}
\usepackage[makeroom]{cancel}
\usepackage[T1]{fontenc}
\usepackage[usenames,dvipsnames]{color}
\usepackage{braket}
\usepackage{bbold}
\usepackage[colorlinks=true,
linkcolor=black,
citecolor=black,
urlcolor =black]{hyperref}

\renewcommand{\vec}[1]{\mathbf{#1}}

\newcommand*{\ketbra}[2]{\ensuremath{\ket{#1}\bra{#2}}}

\newcommand*{\tr}[2][]{\ensuremath{\textrm{Tr}_{#1}\left[ #2 \right]}}
\newcommand{\id}{\ensuremath{{\mathbb{1}}}}
\DeclareMathAlphabet{\mathscr}{OT1}{pzc}{m}{it} 
\DeclareMathOperator{\wg}{Wg}
\graphicspath{{Figures/}}

\begin{document}
	\title[]{Statistical correlations between locally randomized measurements:  a  toolbox for probing entanglement in many-body quantum states}
		\author{A.~Elben}
			\email{andreas.elben@uibk.ac.at}
	\affiliation{Center for Quantum Physics,
		University of Innsbruck, Innsbruck A-6020, Austria}
	\affiliation{Institute for Quantum Optics and Quantum Information, Austrian Academy of Sciences, Innsbruck A-6020,
		Austria}
	\author{B.~Vermersch}
	\affiliation{Center for Quantum Physics,
		University of Innsbruck, Innsbruck A-6020, Austria}
	\affiliation{Institute for Quantum Optics and Quantum Information, Austrian Academy of Sciences, Innsbruck A-6020,
		Austria}
		\author{C.~F.~Roos}
	\affiliation{Center for Quantum Physics,
		University of Innsbruck, Innsbruck A-6020, Austria}
	\affiliation{Institute for Quantum Optics and Quantum Information, Austrian Academy of Sciences, Innsbruck A-6020,
		Austria}
	\author{P.~Zoller}
	\affiliation{Center for Quantum Physics,
		University of Innsbruck, Innsbruck A-6020, Austria}
	\affiliation{Institute for Quantum Optics and Quantum Information, Austrian Academy of Sciences, Innsbruck A-6020,
		Austria}
	
	\begin{abstract}
	We develop a general theoretical framework for measurement protocols employing statistical correlations of randomized measurements. We focus on \emph{locally} randomized measurements implemented with \emph{local} random unitaries in quantum lattice models.  In particular, we discuss the  theoretical details underlying the recent measurement of the second R\'{e}nyi entropy of highly mixed quantum states consisting of up to $10$ qubits in a trapped-ion quantum simulator \href{https://science.sciencemag.org/content/364/6437/260}{[Brydges et al., Science 364, 260 (2019)]}. We generalize the protocol to access the overlap of  quantum states, prepared sequentially in one experiment. Furthermore, we discuss  proposals for quantum state tomography based on randomized measurements within our framework and the respective scaling of statistical errors with system size.
	\end{abstract}
	\maketitle

\maketitle

\section{Introduction}

The  development of intermediate- and large scale quantum simulators \cite{Preskill:2018}, consisting of tens of individually controlled quantum particles, requires  new tools to probe and verify complex many-body quantum systems \cite{Bloch:2012,Blatt2012,Browaeys:2016,Gambetta:2017}. 
 A key feature of  composite quantum systems, in particular quantum lattice models, is bipartite entanglement which can be accessed by the measurement of R\'{e}nyi entropies \cite{Horodecki:2009}. In spin models with a few degrees  of freedom, R\'{e}nyi entropies   can be determined from  tomographic reconstruction of the quantum state of interest \cite{Haeffner2005,Gross:2010,Lanyon:2017, Torlai:2018}. In systems realizing one-dimensional Bose Hubbard models, the measurement of the second R\'{e}nyi entropy has been demonstrated in remarkable experiments  \cite{Islam:2015, Kaufman:2016}. Here, two identical copies of the quantum state have been prepared and the second R\'{e}nyi entropy has been determined from an interference experiment \cite{Ekert:2002,Daley2012,Pichler:2013}.

In Ref.~\cite{Brydges2018}
we have demonstrated  in a theory-experiment collaboration a  new protocol    in which the second order R\'{e}nyi entropy is inferred from statistical correlations of locally randomized measurements \cite{Elben:2018}.  Here, a spin model in a trapped-ion quantum simulator was realized and the generation of entanglement during quench dynamics was monitored.
The  measurement protocol to access the second R\'{e}nyi entropy was based on only  \emph{local operations on individual spins and a single instance of the quantum state}. It required, albeit an exponential scaling with the number of degrees of freedom, a significantly lower number of measurements than standard  quantum state tomography \cite{Brydges2018}.  This measurement protocol is thus immediately applicable  in a broad class of quantum simulators, realizing spin models, with single site readout and control. In particular, we have in mind systems based on trapped-ions \cite{Zhang:2017a,Brydges2018}, Rydberg atoms \cite{Zeiher2017,Barredo2018,Guardado2018,Keesling2018} and superconducting qubits \cite{Blumoff2016,Barends2016,Otterbach2017,Gong2018} in arbitrary spatial dimensions. Moreover, our protocol can straightforwardly be  applied to extended systems such as quantum networks.  {In this context, it has also been proposed use randomized measurements to detect genuine multipartite correlations \cite{Ketterer2018}.}

The key ingredient for the protocols described in this paper are statistical correlations between  randomized measurements.  {The goal of this article is to} develop a general mathematical formalism to evaluate such  correlations and, equipped with this toolbox,  elaborate on the theoretical details behind the protocol realized in Ref.~\cite{Brydges2018}. 

In general, a random measurement on a (reduced) quantum state $\rho$ is performed by the application of  a random unitary $U$, sampled from an appropriate ensemble (see below), and the subsequent measurement of the expectation value $\langle O \rangle_U =\tr[]{U\rho U^\dagger O}$ of a fixed observable $O$.
In this paper, we focus on spin models and \emph{locally} randomized measurements where the random unitaries are of the form $U=\otimes_{i} U_i$ with the $U_i$ independent random spin rotations sampled from unitary designs  (in particular the circular unitary ensemble (CUE)) \cite{Elben:2018,Brydges2018}. Furthermore, we provide an in-depth comparison with protocols based on  \emph{globally} randomized measurements \cite{vanEnk:2012,Elben:2018} where global random unitaries $U$ are sampled from  unitary designs defined on the entire Hilbert space. These global random unitaries can be generated in interacting quantum lattice models with engineered disorder using random quenches \cite{Nakata:2017,Elben:2018,Vermersch:2018} and the corresponding protocols are in particular  relevant for atomic Hubbard models \cite{Elben:2018,Vermersch:2018}.

In the second part, we extend the formalism to derive a protocol to measure  the overlap $\tr[]{\rho \rho'}$ of two  states $\rho$ and $\rho'$ which are prepared sequentially  in one experiment. This allows in particular to directly measure the many-body Loschmidt echo \cite{Goussev2012}, without implementing  time-reversed operations. 
Finally, we discuss within our formalism a proposal of Ohliger et al.~\cite{Ohliger2013} to use randomized measurements, implemented with global random unitaries,  to perform full quantum state tomography in atomic Hubbard models. We generalize this protocol to local random unitaries available in spin models and investigate in detail numerically how the required number of measurements to  reconstruct the density matrix $\rho$ up to a fixed error scales with system size.

\section{Measurement of the second R\'{e}nyi entropy}

In this section, we discuss the measurement of the second R\'{e}nyi entropy in quantum lattice models. After a short review, we focus on protocols utilizing randomized measurements, give explicit  recipes and discuss  examples.
We consider a lattice system $\mathcal{S}$ described by a quantum state $\rho$.  The second R\'{e}nyi entropy $S_2(\rho_A)$ of the reduced density matrix $\rho_A= \tr[\mathcal{S} \backslash A]{\rho}$ of an arbitrary subsystem $A\subseteq  \mathcal{S}$ consisting of $N_A$ sites is defined as
\begin{align}
	S_2(\rho_A) = - \log_2 \tr[]{\rho_A^2}.
\end{align}
Using $S_2(\rho_A)$, one shows that bipartite entanglement exists between two disjoint subsystems $A$ and $B$ of $\mathcal{S}$ with reduced density matrices $\rho_A = \tr[\mathcal{S} \backslash A]{\rho}$ and  $\rho_B = \tr[\mathcal{S} \backslash B]{\rho}$  \footnote{We note that this criterium is sufficient but not necessary.} if 
\begin{align}
\text{ {or}} \quad \begin{split}
S_2\left(\rho_A \right)& >S_2\left(\rho_{A \cup B}\right) \\
S_2\left(\rho_B\right)&> S_2\left(\rho_{A \cup B}\right) ,
\end{split}
\end{align}
where $\rho_{A \cup B}=\tr[\mathcal{S} \backslash A \cup B]{\rho}$ is the reduced density matrix of $A \cup B$ \cite{Horodecki:2009}.

To  measure $S_2(\rho_A)$, i.e.\  the purity $\tr[]{\rho^2_A}$ of a (reduced) density matrix $\rho_A$, various protocols have been proposed and realized which we shortly review in the following:
 A first option realized in spin models is to perform full quantum state tomography of $\rho_A$ \cite{Haeffner2005,Gross:2010,Lanyon:2017, Torlai:2018}.
 However, due to the exponential scaling of the number measurement settings, at least $d^{2N_A}$ for $N_A$ spins and standard tomography \cite{Gross:2010}, this approach is limited to systems with a few degrees of freedom \cite{Haeffner2005}.  On the contrary, recent efficient tomographic methods require a specific structure of the state of interest \cite{Lanyon:2017,Torlai:2018}.

A second class of protocols \cite{Daley2012,Pichler:2013} is based on noting that the purity
\begin{align}
	\tr[]{\rho_A^2} = \tr[]{\mathbb{S}\rho_A \otimes \rho_A}
\end{align}
can be obtained from measuring the expectation value of  the swap operator $\mathbb{S}$ acting on two copies of a quantum state $\rho_A$ \cite{Ekert:2002,Horodecki:2003}.  Here, $\mathbb{S}$ is defined by $\mathbb{S} \ket{s_A} \otimes \ket{s_A'} =\ket{s_A'} \otimes \ket{s_A}  $ for any two states $\ket{s_A}, \ket{s_A'}$.  Prerequisite for these protocols is  thus the experimental ability to create two identical copies  $\rho_A \otimes \rho_A$ and to perform joint operations on them to measure $ \tr[]{\mathbb{S}\rho_A \otimes \rho_A}=\tr[]{\rho_A^2}$. 
Despite the experimental complexity of this task,  the purity of a quantum state of up to six particles, realizing a one-dimensional Bose-Hubbard model, has been measured in remarkable experiments \cite{Islam:2015,Kaufman:2016}. Recently, the method has been also transferred to trapped-ion quantum simulator, and applied to a one-dimensional system of five qubits \cite{Linke:2017}. Creating identical copies of a quantum state in larger systems and  higher spatial dimensions remains however a significant technological  challenge.  

Only a single instance of a quantum state is required in a third class of protocols \cite{vanEnk:2012,Elben:2018,Vermersch:2018,Brydges2018}, utilizing statistical correlations of randomized measurements,  which we discuss in the remainder of this paper.

\subsection{Second R\'{e}nyi entropy from statistical correlations of randomized measurements}
\label{sec:prot}

 \begin{figure}
	\includegraphics[width=\linewidth]{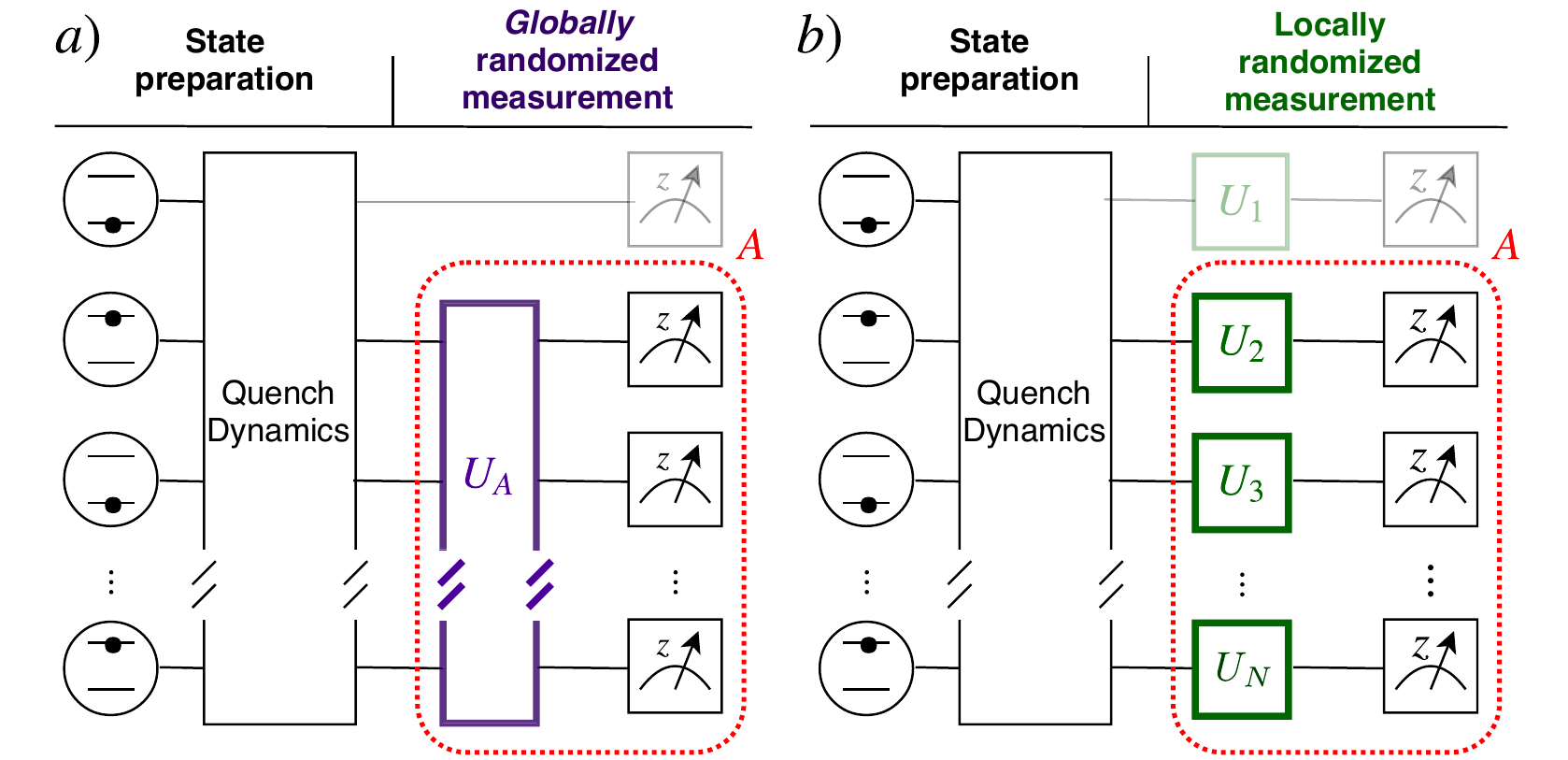}
	\caption{\emph{Measurement of the second R\'{e}nyi entropy using statistical correlations between randomized measurements.}  A  quantum state $\rho$ of interest is prepared for instance via quench dynamics (see also Ref.~\cite{Brydges2018}). A  randomized measurement on a subsystem $A$ is performed by the application of a) a global random unitary $U_A$ from a unitary $2$-design on the entire Hilbert space  or b) a product of local random unitaries $U_A=\bigotimes_{i\in A} U_i$  sampled independently from a unitary $2$-design on $\mathscr h$. Subsequently, a measurement in the computational basis is performed. From statistical correlations of the outcomes of such randomized measurements, the purity $\tr[]{\rho_A^2}$ of the reduced density matrix $\rho_A$ is  estimated (see text).}
	\label{fig:prot}
\end{figure}

In this subsection, we describe the protocol to estimate the purity $\tr[]{\rho^2_A}$ of  a (reduced) density matrix $\rho_A$ from statistical correlation of randomized measurements. While protocols based on global random unitaries are also applicable to atomic Hubbard models \cite{Elben:2018,Vermersch:2018}, we focus in this paper on spin models consisting of $N$ spins with local Hilbert space $\mathscr{h}$ of dimension $d$ (i.e.~$N$ qudits). Here,  in addition, experimentally simpler local random unitaries are available (see below).
A schematic view of  the experimental sequence  we have in mind is displayed in Fig.~\ref{fig:prot}  (see also Ref.~\cite{Brydges2018,Elben:2018}). 
An (entangled) quantum state of interest  $\rho$ is for instance prepared  via quench dynamics originating from a simple initial state. The experimental protocol  to measure the purity of the reduced density matrix $\rho_A$ of a subsystem $A$ of $N_A$ qudits consists then  of several steps. First, one applies to $\rho_A$ a random unitary $U_A$. This can (i) either be a \emph{global} random unitary sampled from a unitary $2$-design \cite{Gross2007} defined on the entire Hilbert space $\mathcal{H}_A=\mathscr{h}^{\otimes N_A}$ with dimension $\mathcal{D}_A=d^{N_A}$ of the subsystem, or (ii)  a \emph{local} unitary of the form 
$
U_a= \bigotimes_{i\in A } U_i
$
where the $U_i$ are sampled independently from a unitary $2$-design defined on the local Hilbert space $\mathscr{h}$. We compare both approaches in detail below. Subsequently, a measurement in the computational basis is performed, which is repeated with the same random unitary $U_A$ to estimate the occupation probabilities $P_U(\vec{s}_A)= \tr[]{U_A\rho U_A^\dagger \ketbra{\vec{s}_A}{\vec{s}_A}}$ of computational basis states $\ket{\vec{s}_A}=\ket{s_1,\dots,s_{N_A}}$ ($s_i=1\dots d$ for $i \in A$). In a second step, this is repeated for many different random unitaries, to estimate the  average over the ensemble of random unitaries.

Given the set of outcome probabilities $P_U(\vec{s}_A)$ for the computational basis states $\vec{s}_A$, one estimates the purity of $\rho_A$ from second-order cross correlations across the random unitary ensemble. For  \emph{global} random unitaries [case~(i)], one finds 
\begin{align}
\tr[]{\rho_A^2}& = \mathcal{D}_A \sum_{\vec{s}_A,\vec{s}'_A} (- \mathcal{D}_A )^{-D_G[\vec{s}_A,\vec{s}_A']}  \; \overline{  P_U(\vec{s}_A) P_U(\vec{s}_A') } \nonumber \\
&=  (\mathcal{D}_A  +1)  \sum_{\vec{s}} \overline{  P_U(\vec{s}_A)^2 } -1,
\label{eq:AkkG}
\end{align}
where  the ‘‘global’’ Hamming distance is  defined as  {$D_G[\vec{s}_A,\vec{s}'_A]=0$} if $\vec{s}_A=\vec{s}_A'$ and  $D_G[\vec{s}_A,\vec{s}_A']=1$ if $\vec{s}_A\neq\vec{s}_A'$. The expression in the second line has first been given in Ref.~\cite{vanEnk:2012}.
If independent \emph{local} random unitaries on individual qudits are used [case~(ii)], the purity is estimated from 
\begin{align}
\tr[]{\rho_A^2} = d^{N_A} \sum_{\vec{s}_A,\vec{s}_A'} (-d)^{-D[\vec{s}_A,\vec{s}_A']}  \; \overline{  P_U(\vec{s}_A) P_U(\vec{s}_A') } ,
\label{eq:Akkpur}
\end{align}
where the Hamming distance $D[\mathbf{s}_A,\mathbf{s}_A']$  between  two  states $\ket{\mathbf{s}_A}=\ket{s_1,\dots,s_{N_A}}$ and $\ket{\mathbf{s}_{A}'}=\ket{s'_1,\dots,s'_{N_A}}$ is defined as  the number of local constituents $i\in A$    where $s_{i} \neq {s}'_{i}$, i.e.\  $D[\mathbf{s}_A,{\mathbf{s}_A}']\equiv   \# \left\{ i \in A  \, | \, s_{i} \neq {s}'_{i} \right\}$. 
 {Equation~\eqref{eq:Akkpur} has first been obtained  in Ref.~\cite{Brydges2018}, and represents an explicit version of the recursive formula given in Ref.~\cite{Elben:2018}. In Ref.~\cite{Brydges2018}, the identity has been established directly using moments of matrix elements of random unitaries. In this paper, we develop a general formalism to evaluate statistical correlations of randomized measurements based on  the Weingarten calculus \cite{Collins2009}. This approach enables a constructive and simple derivation of Eq.~\eqref{eq:Akkpur} and, in addition, it simplifies the development of new protocols (see below). }

An intriguing connection to the previous works realizing  the swap operator on two \emph{physical copies}  \cite{Daley2012,Pichler:2013,Islam:2015,Kaufman:2016}  can be seen as follows: We can rewrite any product  of outcome probabilities $P_U(\vec{s}_A) P_U(\vec{s}_A') = \tr[\mathcal{H}^{\otimes 2}_A ]{U_A^{\otimes 2} \rho_A^{\otimes 2}U_A^{\dagger \otimes2 }  \ketbra{\vec{s}_A}{\vec{s}_A} \otimes \ketbra{\vec{s}_A'}{\vec{s}_A'}}$ as the expectation value of an operator $\ketbra{\vec{s}_A}{\vec{s}_A} \otimes \ketbra{\vec{s}_A'}{\vec{s}_A'}$ on the doubled Hilbert space $\mathcal{H}^{\otimes 2}_A $. Using this, we  can  intuitively understand  the ensemble average over second order cross correlations taken in Eqs.~\eqref{eq:AkkG} and \eqref{eq:Akkpur} as an effective construction of the swap operator on two \emph{virtual copies} of $\rho_A$ (see for details Sec.~\ref{sec:rnduni}). 

The approaches with  global and local random unitaries differ in various aspects:
First, the implementation of {global random unitaries} from a unitary $2$-design acting on the entire many-body quantum state $\rho_A$ requires interactions between the particles.  It has been  proposed to prepare them efficiently in quantum circuits using   (random) entangling gates \cite{Dankert2009} or in generic interacting many-body systems using  time evolution subject to random quenches  \cite{Ohliger2013,Nakata:2017,Elben:2018,Vermersch:2018}.  On the contrary, {local random unitaries}, available in spin models, are single ``qudit'' operations (random spin rotations) which have been demonstrated with high fidelity and repetition rates \cite{Brydges2018}.
Second, the protocol utilizing local random unitaries allows, from a single experimental dataset obtained from randomized measurements on the subsystem $A$,  to estimate the purity $\tr[]{\rho_{A'}^2}$ of the reduced density matrix $\rho_{A'}$ of any  subsystem $A'\subseteq A$. To this end,   Eq.~\eqref{eq:Akkpur} is evaluated with occupation probabilities $P_U(\mathbf{s}_{A'})$ of  states $\ket{\vec{s}_{A'}}$ of the logical basis of $A'$.  
Third, the two protocols differ in their sensitivity to statistical errors. In an experiment, statistical errors of the estimated purity arise from a finite number $N_U$ of applied unitaries and a finite number $N_M$ of measurements  per random unitary (projection noise). The total number of measurements $N_MN_U$ scales exponentially  with the number $N_A$ of degrees of freedom in the subsystem $A$, with exponents significantly smaller than in full quantum state tomography \cite{Brydges2018,Elben:2018}. As we discuss in detail below, the protocol utilizing  global unitaries is, for pure product states,  favorable in terms of statistical errors.

To prove Eqs.~\eqref{eq:AkkG} and \eqref{eq:Akkpur}, we introduce results of the theory of  random unitaries   in Sec.~\ref{sec:rnduni}. The proof follows then in Sec.~\ref{sec:proof}.

  \subsection{Illustrative examples}

  In the remainder of this section, we illustrate  Eqs.\ \eqref{eq:AkkG} and \eqref{eq:Akkpur} using simple examples.

  \subsubsection{Single qubit}

\begin{figure}
 	\centering
 	\includegraphics[width=0.5\linewidth]{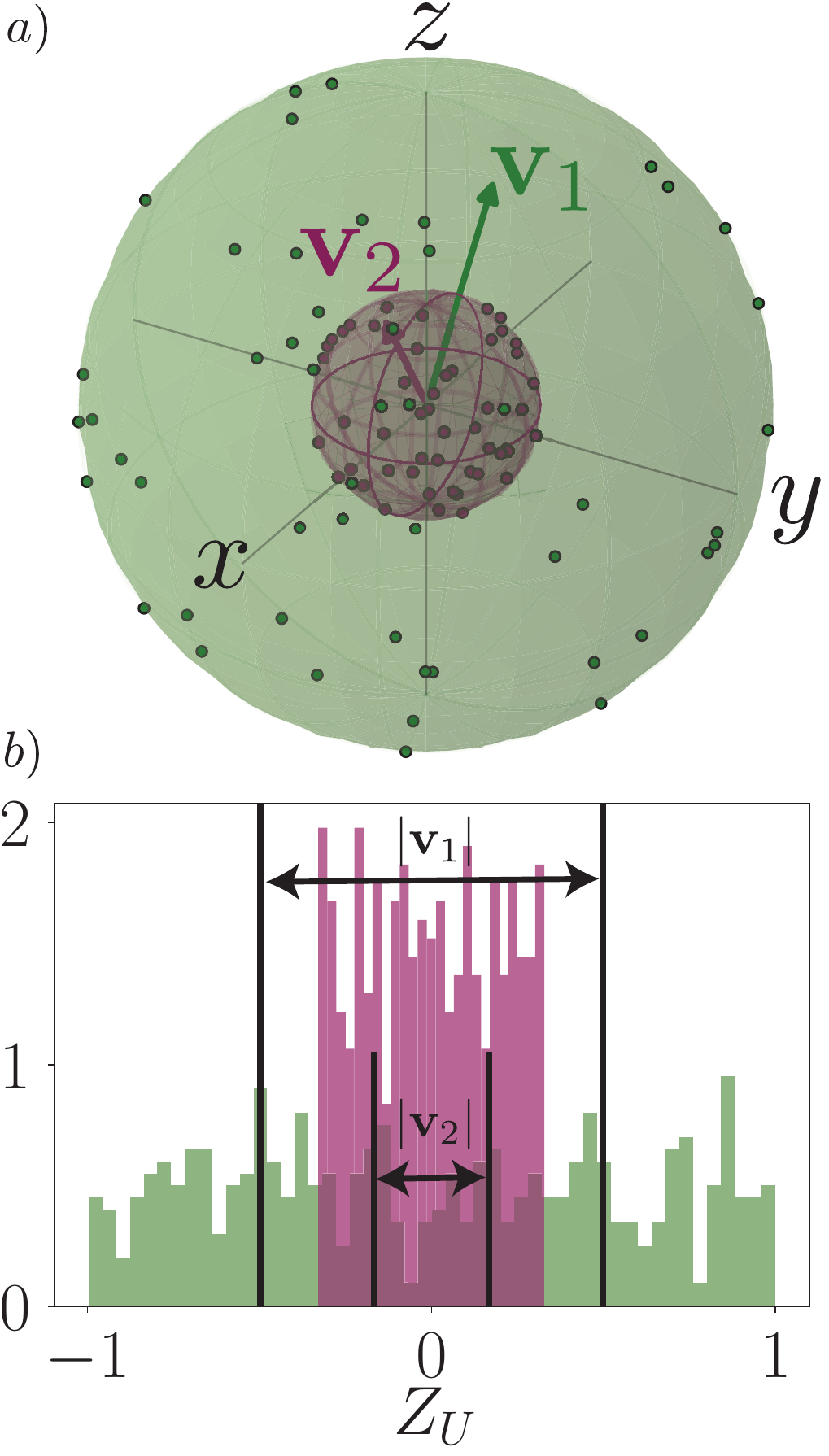}
 	\caption{a) Graphical visualization of a pure $\rho_1$ (green) and a mixed  $\rho_1$ (purple) singe qubit state with Bloch vectors (arrows)  $\mathbf{v}_1$ and  $\mathbf{v}_2$, respectively. Points correspond to 50 randomly rotated states, generated via the application of random unitaries sampled from the CUE \cite{Mezzadri:2007} to $\rho_1$ and $\rho_2$, respectively. b) Histogram  of the random variable $Z_U=\tr[]{U\rho U^\dagger \sigma_z}$ for pure  $\rho_1$ (green) and mixed $\rho_2$ (purple) state, the indicated standard deviation (multiplied with a factor $\sqrt{3}$) corresponds to the length of the  Bloch vectors.}
 	\label{fig:Fig1Qubit}
 \end{figure}

 The density matrix $\rho=\frac{1}{2} \left( \id_2 + \vec{v} \cdot \vec{\sigma} \right)$ of a single qubit is conveniently represented on the Bloch sphere, with the real Bloch vector $(\vec{v})_i=\tr[]{\rho \sigma_i}$ and $\mathbb{\sigma}=(\sigma_x,\sigma_y,\sigma_z)$ the Pauli matrices. The purity $\tr[]{\rho^2}=\frac{1}{2} \left(1+ |\vec{v}|^2 \right)$ is thus fully determined by the length of the Bloch vector $|\vec{v}|$. Our approach to estimate the length of the Bloch vector consists  in applying a random unitary $U$ to the state $\rho$ and measuring the difference of occupation $Z_U=P_U(\uparrow)- P_U(\downarrow)= \tr[]{U\rho U^\dagger \sigma_z}=\left(Q_U \vec{v}\right)_3$ of the computational basis states $\ket{\uparrow}, \ket{\downarrow}$.  Here, $Q_U$ is the unique rotation matrix corresponding to the unitary $U$, i.e.\  $U \vec{v} \cdot \sigma U^\dagger = \left(Q_U \vec{v}\right) \cdot \sigma$ for all Bloch vectors $\vec{v}$ \cite{Gamel2016}. This is now repeated with different random unitaries, to sample the distribution of $Z_U$ across the circular unitary ensemble.  {Since the random variable $Z_U=\left(Q_U \vec{v}\right)_3$ describes the $z$-component of the rotated Bloch vector $Q_U \vec{v}$ subject to random rotations $Q_U$, it is intuitively apparent that its range contains information about the length of $\mathbf{v}$ (see also Fig.~\ref{fig:Fig1Qubit}). }
 
 Formally, we note that moments $\overline{Z_U^n}$ ($n \in \mathbb{N}$) of the random variable $Z_U$ are invariant under  unitary transformations due to the invariance properties of  the Haar measure [see below, Eq.~\eqref{eq:Haar}].  Thus, they must be determined by properties of the Bloch vector $\vec{v}$ which are invariant under arbitrary rotations. The squared length $|\vec{v}|^2$ is its unique second order invariant, and thus we conclude $\overline{   Z_U^2} = \overline{ \left(Q_U \vec{v}\right)^2_3} \sim |\vec{v}|^2  $.  Indeed, using the $n$-design properties, we find that $Z_U$ is uniformly distributed, with zero mean and variance $\overline{   Z_U^2} =|\vec{v}|^2/3$, such that $\tr[]{\rho^2}=\frac{1}{2} \left(1+ 3 \overline{(P_U(\uparrow)- P_U(\downarrow))^2} \right)$. Inserting that $1=(P_U(\uparrow)+ P_U(\downarrow))^2$ we can bring this into a more symmetric form to arrive at
 \begin{align}
 \tr[]{\rho^2} = 2 \left(  \overline{  P_U(\uparrow)^2 + P_U(\downarrow)^2 -  P_U(\uparrow)P_U(\downarrow)} \right)
 \end{align}
 which corresponds to Eq.~\eqref{eq:AkkG} [Eq.~\eqref{eq:Akkpur}] for the special case of $\mathcal{D}_A=2$  [$d=2$ and $N_A=1$].

 \subsubsection{Two qubits}
 
 We consider now the case of two qubits where randomized measurements are implemented using independent local random unitaries $U=U_1\otimes U_2$. 
 Generalizing the single-qubit case, the two-qubit system is conveniently represented in the basis of Pauli strings
 \begin{align}
 \rho= \frac{1}{4}\sum_{\mu , \nu = 0} ^3  r_{\mu \nu} \;\sigma_\mu \otimes  \sigma_\nu  ,
 \end{align}
 where $\sigma_0=\id_2$ and $\sigma_i$  the Pauli matrices ($1\leq i \leq 3$). The real coefficients $r_{\mu \nu}=\tr[]{\rho \sigma_\mu \otimes \sigma_\nu}$  constitute the Bloch matrix $\mathbf{r}$ (generalizing the Bloch vector)
 \begin{equation}
 r=\left(\begin{array}{c|ccc}
 1 & r_{01} & r_{02} & r_{03}\\
 \hline r_{10} & r_{11} & r_{12} & r_{13}\\
 r_{20} & r_{21} & r_{22} & r_{23}\\
 r_{30} & r_{31} & r_{32} & r_{33}
 \end{array}\right)\equiv \left(\begin{array}{cc}1 & \vec{w}^{T}\\
 \vec{v} & R
 \end{array}\right),\label{eq:p2uvP}
 \end{equation}
 where $v_{i}=r_{i0}$, $w_{j}=r_{0j}$, and $R_{ij}=r_{ij}$. The vectors $\vec{v}$ and $\vec{w}$ are the Bloch vectors of the reduced density matrices $\rho_1 =\tr[2]{\rho}$  and $\rho_2 =\tr[1]{\rho}$ of the individual qubits, respectively. The matrix $R$ quantifies the correlations between the two qubits \cite{Gamel2016}. Using these definitions, the purity of the density matrix $\rho$ is then given by
 \begin{equation}
 \tr[]{\rho^2} = \frac{1}{4} \left( 1 + |\mathbf{v}|^2 +|\mathbf{w}|^2+ ||R||^2  \right) ,
 \end{equation}
 where $||R||^2= \tr[]{R^\dagger R}$.
 
 \begin{figure}
 	\centering
 	\includegraphics[width=0.95\linewidth]{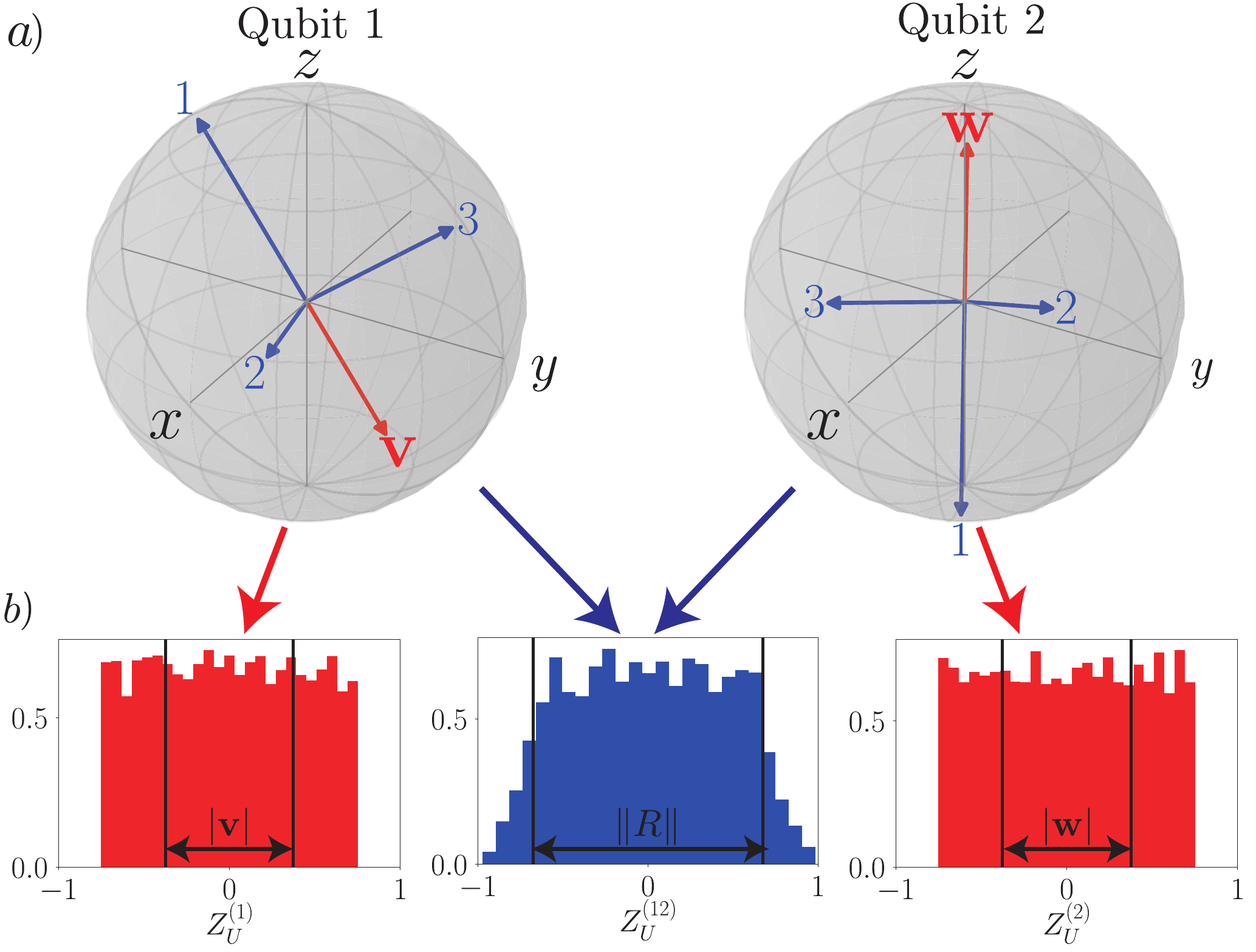}
 	\caption{a) Graphical visualization of a pure, random (entangled) two-qubit  state. The red arrows visualize the Bloch vectors $\vec{v}$ and $\vec{w}$ of reduced (mixed) single qubit states. Blue arrows correspond to the  left   $\sqrt{\gamma_i} \vec{m}_i$ and right   $\sqrt{\gamma_i} \vec{n}_i$ singular  vectors, rescaled with singular values $\gamma_i$,   of the singular value decomposition of $R$ ($i=1,\dots,3$) \cite{Gamel2016}.  The measurement of $\sigma_z \otimes \sigma_z$ after the application of a unitary  $U_1 \otimes U_2$  (rotations $Q_1$ and $Q_2$ of qubit 1 and 2)  is visualized as a projection of the singular vectors onto the $z$-axis, its expectation value is given as $\sum_i \gamma_i  (Q_1 \vec{m}_i)_3   (Q_2 \vec{n}_i)_3 $.
 		b)   Histograms of random variables  $Z_U^{(1)}=\tr[]{U \rho U^\dagger \sigma_z \otimes \mathbb{1}_2} $, $Z_U^{(2)}=\tr[]{U \rho U^\dagger  \mathbb{1}_2 \otimes \sigma_z} $ and  $Z_U^{(12)}=\tr[]{U \rho U^\dagger  \sigma_z \otimes \sigma_z} $ generated using random unitaries of the form $U=U_1 \otimes U_2$.  The  standard deviation (multiplied with a factor $\sqrt{3}$ for left and right and $3$ for the middle panel) corresponds to the length of the Bloch vectors $|\vec{v}|$ and $|\vec{w}|$ (left and right) and the Hilbert-Schmidt norm $\|R\|$ of the correlation matrix $R$ (middle), see text. }
 \end{figure}
 
 Our aim is to estimate the purity $\tr[]{\rho^2}$ using  random unitaries of the $U_1 \otimes U_2$ and (collective) measurements in the computational basis $\ket{\uparrow \uparrow},\ket{\uparrow \downarrow},\ket{\downarrow \uparrow},\ket{\downarrow \downarrow}$.  First, we note that under  unitary transformation $\rho \rightarrow U_1 \otimes U_2 \rho  U_1^\dagger \otimes U_2^\dagger$, the individual elements of the Bloch matrix transform as 
 \begin{align}
 \begin{split}
 \vec{v} &\rightarrow \vec{v}' = Q_1 \vec{v}\\
 \vec{w} &\rightarrow \vec{w}' = Q_2 \vec{w}\\
 {R} &\rightarrow {R}' = Q_1 R Q_2^\dagger,
 \end{split}
 \label{eq:transfo2qubit}
 \end{align}
 there $Q_1$ ($Q_2$) is the unique rotation matrix corresponding to $U_1$ ($U_2$) \cite{Gamel2016}.  Thus  $|\mathbf{v}|^2$ and $|\mathbf{w}|^2$ can be estimated  from single qubit measurements, i.e.\ from the variances of the distributions \begin{align}
Z^{(1)}_U&=P^{(1)}_U(\uparrow)-P^{(1)}_U(\downarrow)\nonumber \\&= P_U(\uparrow\uparrow)+P_U(\uparrow\downarrow)-P_U(\downarrow\uparrow)-P_U(\downarrow\downarrow)
 \end{align} and $Z^{(2)}_U$, respectively.  
 The unique element of the (transformed) correlation matrix accessible from measurements solely in the computational basis is \begin{align}
Z^{(1,2)}_U &=\tr[]{ U_1 \otimes U_2 \rho  U_1^\dagger \otimes U_2^\dagger \sigma_z\otimes \sigma_z} \nonumber \\
&=P_U(\uparrow\uparrow)-P_U(\uparrow\downarrow)-P_U(\downarrow\uparrow)+P_U(\downarrow\downarrow)\nonumber \\&= \left( Q_1 R Q_2^\dagger\right)_{33} .
 \end{align}
 Again, due to the Haar average, moments of the random variable $Z^{(1,2)}_U$ must be invariant transformations \eqref{eq:transfo2qubit} with arbitrary rotations $Q_1$ and $Q_2$, and a  second order invariant of this type is the matrix norm $||R||_2$.  Indeed, we find $\overline{  \left( Z^{(1,2)}_U \right)^2 }= ||R||_2/9  $.
 Thus we obtain, 
 \begin{align}
 \tr[]{\rho^2}&= \frac{1}{4} \left( 1 +  3 \overline{ \left(Z^{(1)}_U \right)^2  } + 3 \overline{ \left(Z^{(2)}_U \right)^2  } +  9 \overline{ \left(Z^{(1,2)}_U \right)^2  } \right) \nonumber \\
 &= 4 \sum_{\stackrel{\vec{s},\vec{s}'=}{ \uparrow\uparrow,\uparrow\downarrow,\downarrow\uparrow,\downarrow\downarrow}}  (-2)^{-D[\vec{s},\vec{s}']}  \overline{P_U(\vec{s})P_U(\vec{s}')},
 \end{align}
 which corresponds to formula \eqref{eq:Akkpur} for the special case of $d=2$ and $N_A=2$. To arrive at the last line, we used that $1=\left(P_U(\uparrow\uparrow)+P_U(\uparrow\downarrow)+P_U(\downarrow\uparrow)+P_U(\downarrow\downarrow) \right)^2$. We note that these arguments generalize  to $N_A$-qubit systems whose density matrix can be parametrized by a rank $N_A$ ``Bloch tensor''.  The second moment of a suitably generalized random variable $Z^{(1,\dots,N_A)}_U = \tr[]{\bigotimes_{i=1}^{N_A} U_i \rho_A \bigotimes_{i=1}^{N_A} U_i^\dagger \sigma_z^{\otimes N_A}}$ is connected to the Hilbert-Schmidt norm $||R^{(N_A)}||^2$ of the rank $N_A$ ``correlation tensor'' $R^{(N_A)}$.

\section{Random unitaries,  diagrammatic calculus and unitary $k$-designs }
\label{sec:rnduni}

In this section, we provide an overview over the mathematical tools necessary to describe randomized measurements based on random unitaries. We consider first a single qudit with Hilbert space $\mathscr h$ of (arbitrary) dimension $d$, which also  corresponds to the case of global random unitaries (with the replacement $\mathscr{h} \rightarrow\mathcal{H}_A $, $d\rightarrow \mathcal{D}_A=d^{N_A}$, see Sec.~\ref{sec:prot}). In this setting, we  introduce the Haar measure, defining the CUE, and derive elementary properties of the unitary twirling channel. Moreover, we introduce a graphical calculus simplifying the calculations. In the last subsection, we extend the framework  to composite systems of many qudits where multiple, independent, random unitaries are applied locally.

\subsection{Haar randomness }

We first provide an overview over Haar random unitaries and introduce the central object of our formalism, the unitary twirling channel. We follow the treatment of Ref.~\cite{Roberts2017}. An orthonormal basis of $\mathscr h$ is denoted with $\left\{\ket{{s}}\right\}$.

The most important ingredient are Haar random unitaries. These are unitary matrices which are distributed according to the probability distribution defined by the Haar measure on the unitary group \footnote{Intuitively, Haar random unitaries are matrices with elements whose real and imaginary parts are independently distributed according to a normal distribution, with additional unitary constraints on the entire matrix.}. Here, the Haar measure is the unique probability measure on the group of unitary matrices $\mathcal{U}(\mathscr{h})$ on $\mathscr{h}$ which is both left- and right-invariant, i.e.\ it satisfies for any function $f$ on $\mathcal{U}(\mathscr{h})$  and any unitary $V\in \mathcal{U}(\mathscr{h})$
\begin{align}
	&\int_{\textrm{Haar}} \!\!\!\!\textrm{d}U =1, \nonumber \\
	& \int_{\textrm{Haar}} \!\!\!\!\textrm{d}U f(VU) = \int_{\textrm{Haar}} \!\!\!\!\textrm{d}U f(UV)  = \int_{\textrm{Haar}}\!\!\!\! \textrm{d}U f(U) \; .
	\label{eq:Haar}
\end{align}
Often,  $\mathcal{U}(\mathscr{h})$  equipped with the Haar measure is also called the \emph{circular unitary ensemble} (CUE) \cite{Haake2010}, and we use $\overline{f(U)}\equiv \int_{}\!\textrm{d}U f(U)$ to denote the ensemble average over the CUE.

In the following, we consider  the $k$-fold copy space $\mathscr{h}^{\otimes k}$, $k \in \mathbb{N}$, to calculate higher order moments of random unitaries.  We note that this is a purely mathematical construction: $k$th-order products of outcome probabilities of randomized measurements can be viewed as expectation value an operator acting on $\mathscr{h}^{\otimes k}$, realizing thus $k$ ``virtual copies''. It is a key property of any measurement protocol presented in this paper that only \emph{a single physical instance} of a quantum state is required in the experiment.  
We define on $\mathscr{h}^{\otimes k}$ a quantum channel, the $k$-fold twirl by
\begin{align}
\Phi^{(k)}(O) =   \int_{\textrm{Haar}} \textrm{d}U  \left(U^\dagger\right)^{\otimes k}   O  U^{\otimes k}  . 
\end{align}
for any operator $O$ on $\mathscr{h}^{\otimes k}$.  As a simple consequence of the invariance of the Haar measure [Eq.~\eqref{eq:Haar}], $\Phi^{(k)}$ forms a projector $\Phi^{(k)}(\Phi^{(k)}(O))=\Phi^{(k)}(O)$. We show in the following that its image is spanned by permutation operators $W_\pi$ , for permutations $\pi=(\pi(1),\dots, \pi(k)) \in \mathcal{S}_k$ with the symmetric group $\mathcal{S}_k$,  which are defined
as
\begin{align}
W_{\pi} \; = \sum_{{s}_1,\dots,{s}_k=1}^{d} \ket{{s}_{\pi(1)}} \cdots\ket{{s}_{\pi(k)}} \bra{{s}_1}\cdots\bra{{s}_k} \; .
\end{align}
These operators permute states between individual copies $W_{\pi}\ket{{s}_{1}} \cdots\ket{{s}_{k}} =\ket{{s}_{\pi(1)}} \cdots\ket{{s}_{\pi(k)}}$.  Using that $[W_\pi, V^{\otimes k}]=0$ for any $\pi \in  \mathcal{S}_k$, it follows directly that $ \Phi^{(k)}(W_\pi)= W_\pi$, i.e. the permutation operators $W_\pi$ are invariant under the projection $\Phi^{(k)}$. Indeed, they span the total image of $\Phi^{(k)}$, which is proved with the Schur Weyl duality \cite{Roberts2017}. Explicitly, one finds
\begin{align}
\Phi^{(k)} (O) = \sum_{\pi,\sigma \in \mathcal{S}_k} C_{\pi,\sigma}   \, \tr[]{W_\sigma O} \, W _\pi,
\label{eq:Phik1}
\end{align}
where the coefficients $C_{\pi,\sigma}=\wg(\pi \sigma ^{-1})$ constitute the real-valued, symmetric Weingarten matrix $C$ determined by the Weingarten function $\wg$  \cite{Collins2009,Puchala2017,Roberts2017}. For $k \leq d$, $C$ is invertible, with inverse $Q\equiv C^{-1}$ and $Q_{\pi,\sigma}=d^{\sharp\textrm{cycles}(\pi\sigma)}$  \cite{Roberts2017}. 
For $k=1$ and $k=2$, we find
\begin{align}
\Phi^{(1)}(O) = \frac{{\id}}{d} \tr[]{O}
\label{eq:phi11}
\end{align}
and
\begin{align}
\begin{split}
\Phi^{(2)} (O) =\frac{1}{d^2-1} &\left( \vphantom{\frac{1}{d}} \id \tr[]{O} + \mathbb{S} \tr[]{ \mathbb{S}O} \right.  \\ & \left.- \frac{1}{d}  \mathbb{S} \tr[]{O} - \frac{1}{d} \id \tr[]{ \mathbb{S}O} \right),
\end{split}
\label{eq:phi21}
\end{align}
with the identity $\mathbb{1}=W_{(1,2)}$  and  $ \mathbb{S} =  W_{(2,1)}=\sum_{{s}_1{s}_2}\ket{{s}_{2}}\ket{{s}_{1}} \bra{{s}_1}\bra{{s}_2} $ being the swap operator.

\subsection{Diagrammatic calculus}
\label{sec:diagram}

\begin{figure}
	\centering
	\includegraphics[width=0.66\linewidth]{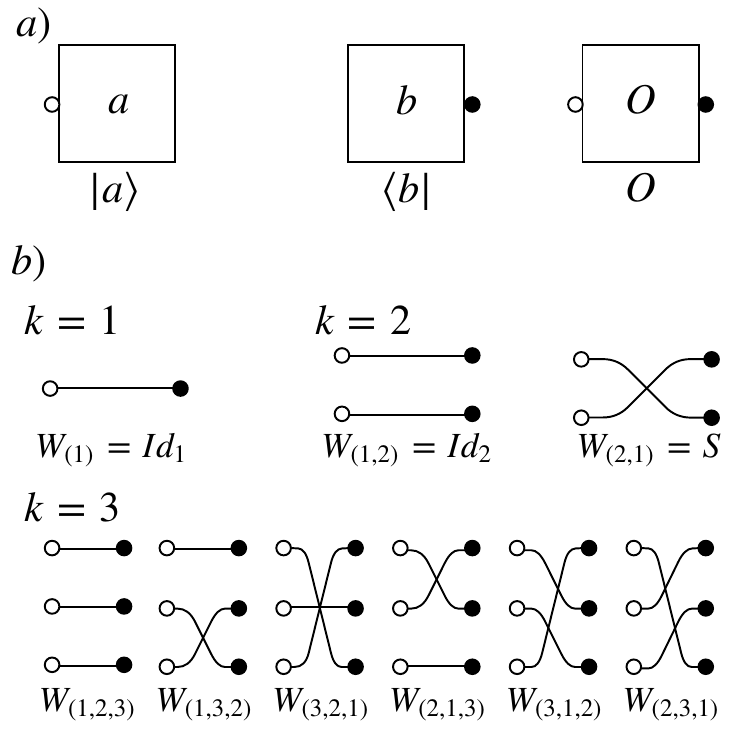}
	\caption{Graphical dictionary: a) Elementary diagrams, describing ket-vectors [(0,1)-tensors], bra-vectors [(1,0)-tensors] and operators [(1,1)-tensors] on a single copy. b) Graphical representation of permutation operators [$(k,k)$-tensors] acting on the $k$-fold copy space.}	
	\label{fig:Dictionary}
\end{figure}

In the previous section, we showed how to evaluate the unitary twirling channel $\Phi^{(k)}$ in terms of permutation operators. Here, we introduce a graphical calculus which enables the evaluation of arbitrary functionals of random unitaries, in particular $\Phi^{(k)}$.  We follow and adapt here the treatment of Ref.~\cite{Collins2009} (see also Ref.~\cite{Brouwer1996} for a similar approach). We first note that, for any $k \in \mathbb{N}$, an operator $O$ acting on $\mathscr{h}^{\otimes k}$ can be viewed as a $(k,k)$-tensor $O \in \mathscr{h}^{*\otimes k} \otimes \mathscr{h}^{\otimes k}$ with $k$ covariant and $k$ contravariant indices. As shown in Fig.~\ref{fig:Dictionary}(a),   we represent tensors in the following as a box with decorations, where the number of empty (filled) symbols corresponds to the number of contravariant  (covariant) indices. Similarly, ket vectors $\ket{a} \in \mathscr{h}$ can be viewed as a $(0,1)$ tensors with a single white decoration, and  bra vectors $\bra{a} \in \mathscr{h}^*$ are $(1,0)$ tensors with a single black decoration. 
As discussed in the previous section, permutation operators are of central importance to evaluate twirling channels. Thus, we introduce a special graphical notation displayed in Fig.~\ref{fig:Dictionary}(b), which corresponds directly to their action on the $k$-fold copy space.

\begin{figure}
	\centering
	\includegraphics[width=\linewidth]{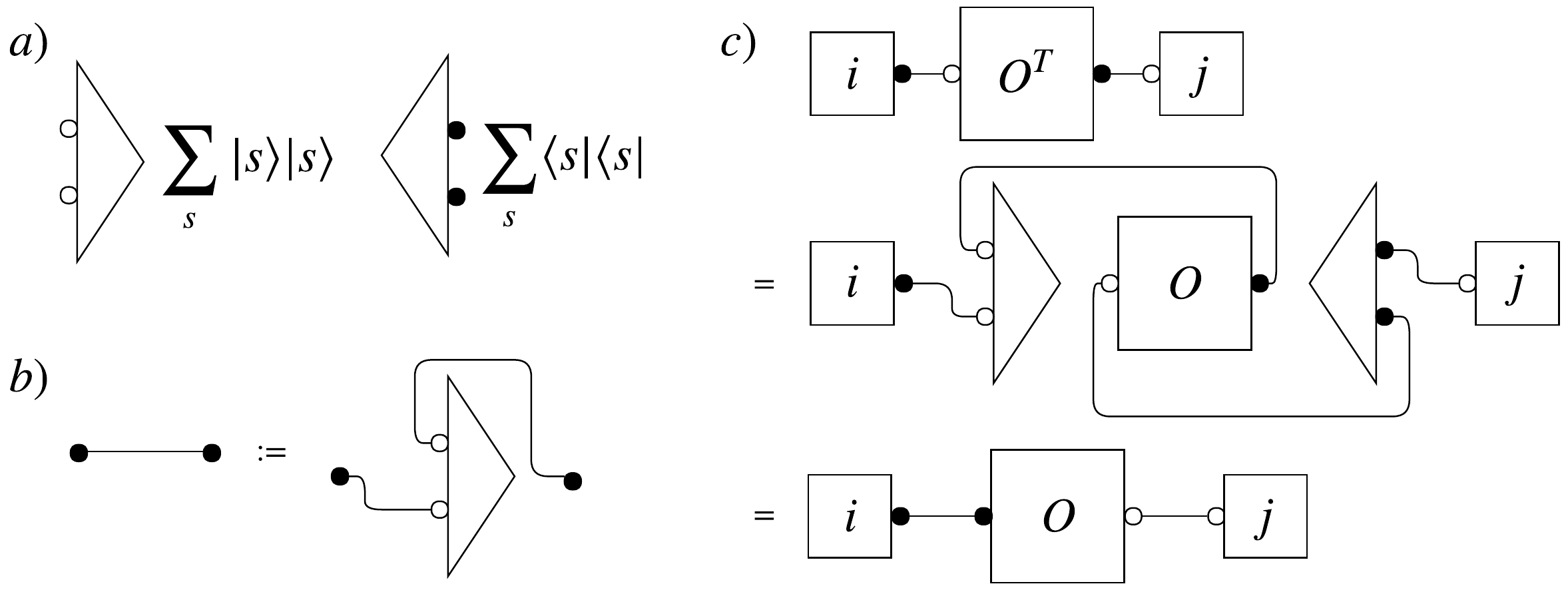}
	\caption{Bell states and the transpose matrix: a) Definition of Bell states [(0,2)- and (2,0)-tensor] on two copies. b) Definition of a diagram connecting two decorations of the same color, in the analogous way the diagram connecting to white decorations is defined. c) Visualization of the transposed matrix using Bell states (see text).}
	\label{fig:BellState}
\end{figure}

In order to evaluate the Haar average of an arbitrary diagram containing random unitaries, it turns out to be useful to consider Bell states $\sum_{i} \ket{i}\ket{i}$ [$\sum_{i} \bra{i}\bra{i}$], which represent special  $(0,2)$ [$(2,0)$] tensors [Fig.~\ref{fig:BellState}(a). Such states allow the definition of graphs connecting two decorations of identical color [Fig.~\ref{fig:BellState}(b)] and, using the identity $(O^T)_{ij}=O_{ji}=\sum_{k,k'} \braket{i|k} \braket{k'|O|k} \braket{k'|j}$, the transposed matrix, given as the tensor with interchanged decorations [Fig.~\ref{fig:BellState}(c)].

\begin{figure}
	\centering
	\includegraphics[width=\linewidth]{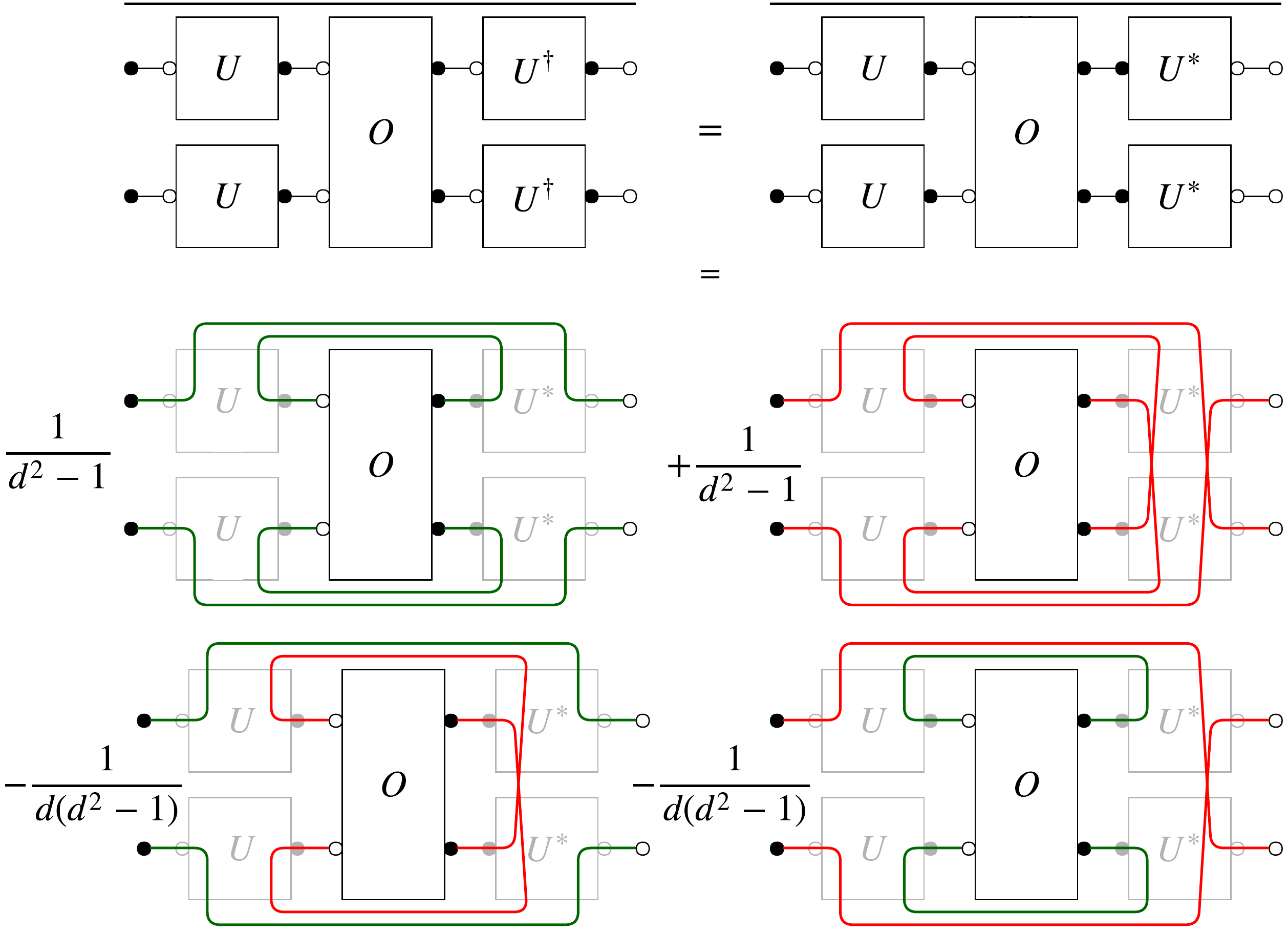}
	\caption{Graphical evaluation of $\Phi^{(2)}$. We evaluate graphical the matrix element $\bra{i}\bra{j} \Phi^{(2)}(O) \ket{i'}\ket{j'}$. For clarity the boxes describing the basis states $\bra{i}\bra{j}$, $\ket{i'}\ket{j'}$ have  been omitted, only their decorations are kept. } 
	\label{fig:graphphi2}
\end{figure}

Equipped with these definitions, we are now in the position to evaluate the ensemble average of an arbitrary diagram. We describe a general procedure \cite{Collins2009} applicable to any diagram. As an explicit example we present  in Fig.~\ref{fig:graphphi2} the evaluation of  $\bra{i}\bra{j} \Phi^{(2)}(U) \ket{i'}\ket{j'}$ \footnote{Note that for clarity, we omitted here the boxes of basis states, and just kept the decorations.}.
The first step is to replace all unitaries $U^T$ with $U$ and $U^\dagger$ with $U^*$ using the Bell states and tensors defined in Fig.~\ref{fig:BellState}. If in the resulting diagram, the number of unitaries $U$ does not equal the number of complex conjugates $U^*$, the diagram evaluates to zero. Otherwise, white decorations of boxes $U$ are connected with white decorations of boxes $U^*$ and black decorations  of boxes $U$  are (independently) connected with black  decorations of boxes $U^*$. Subsequently decorations and boxes of the random unitaries are removed. Given  $k$ boxes $U$, there exist $(k!)^2$ possible ways to draw these connections.
The ensemble average is obtained as a weighted sum of all resulting diagrams, with coefficients determined by the Weingarten matrix. To calculate these coefficients, one labels both the boxes $U$ and boxes $U^*$ with arbitrary integers $1,\dots k$. Each possibility to connect white  (black) decorations is now described by a permutation $\alpha \in S_k$ ($\beta\in S_k$): If the white decoration of $U$-box $i$ is connected to the white decoration of $U^*$-box $j$, then $\alpha(i)\equiv j$. The coefficient of a diagram is obtained as $\wg(\alpha \beta^{-1})$. 
In the case of  Fig.~\ref{fig:graphphi2}, there exist four possibilities, and the resulting sum of diagrams is the graphical representation of  Eq.~\eqref{eq:phi21}.

\subsection{Unitary $k$-designs}

In the previous sections, we considered Haar random unitaries drawn from the circular ensemble for which Eq.~\eqref{eq:Phik1} holds for arbitrary $k\in \mathbb{N}$. In applications, one is however typically interested in moments of random unitaries up to a finite (small) number $k$ \cite{Nakata:2017}, for instance  $k=2$ for the estimation of second order R\'{e}nyi entropy. Since the preparation of Haar random unitaries using, for instance, random quantum circuits requires  an amount of resources scaling exponentially with system size \cite{Nakata:2017}, simpler ensembles, unitary $k$-designs, have been introduced \cite{Gross2007,Roy2009,Dankert2009}. These ensembles approximate Haar random unitaries in the sense that up to  $k$-th order moments are identical, i.e.~Eq.~\eqref{eq:Phik1} holds, for an arbitrary operator $O$, up to a finite, fixed $k$ \footnote{ Loosely speaking, up to the $k$th moment, $k$-designs are as random as Haar random unitaries.}.
To define unitary $k$-designs formally,   we introduce the  $k$-fold twirl with respect to a continuous ensemble $\mathcal{E}$ of unitary operators by
\begin{align}
\Phi^{(k)}_\mathcal{E} (O) =   \int_{\mathcal{E}} \textrm{d}U \,  \left(U^\dagger\right)^{\otimes k}   O U^{\otimes k} ,
\end{align}
and for a discrete ensemble $\mathcal{E}$  with cardinality $|\mathcal{E}|$  by
\begin{align}
\Phi^{(k)}_\mathcal{E} (O) = \frac{1}{|\mathcal{E}|}   \sum_{U \in \mathcal{E}}  \left(U^\dagger\right)^{\otimes k}   O   U^{\otimes k}   . 
\end{align}
One says that $\mathcal{E}$ forms an \emph{unitary $k$-design} if and only if $\Phi^{(k)}_\mathcal{E} = \Phi^{(k)} $ and that $\mathcal{E}$ forms an \emph{ $\epsilon$-approximate $k$-design} if and only if $|| \Phi^{(k)}_\mathcal{E}  -\Phi^{(k)}  ||_\diamond < \epsilon$ \cite{Dankert2009}.  It follows directly that any $k$-design is also an $k'$ design, for any $k'<k$. A prime example of an exact  $3$-design is the Clifford group \cite{Roy2009}. Importantly,  $\epsilon$-approximate $k$-design can be prepared efficiently in local random quantum circuits \cite{Dankert2009}, and generic interacting  quantum simulators \cite[]{Nakata:2017,Elben:2018,Vermersch:2018}.

\subsection{Composite systems}

In this section, we generalize our treatment to composite systems consisting of ${N_A}$ qudits with Hilbert space $\mathcal{H}=\mathscr{h}^{\otimes N_A}$. For simplicity of notation we drop the subscript $A$. The basis $\{\ket{\vec{s}}\}$ denotes a product basis $\ket{\vec{s}}=\otimes_{i=1}^N\ket{s_i}$ for all $\vec{s}=(s_1,\dots,s_N)$.  We consider random unitaries of the form $U=\bigotimes_{i=1}^N U_i$ where the $U_i$ ($i \in \{1,\dots,N\}$), acting on the individual qudits,   are sampled independently from  the $\text{CUE}(\mathscr{h})$ (a unitary $k$-design) defined on the local Hilbert space $\mathscr{h}$. We define a  $k$-fold local twirling channel  by
\begin{align}
\Phi^{(k)}_N (O) \equiv   \overline{ \left( U^{\otimes k} \right) ^\dagger   O U^{\otimes k} } ,
\end{align}
where $\overline{  \vphantom{O} \dots } $ denotes in this context the ensemble average over random unitaries of the form $U=\bigotimes_{i=1}^N U_i$. Generalizing Eq.~\eqref{eq:Phik1}, we find
\begin{align}
\Phi^{(k)}_N (O) = \sum_{\pi,\sigma \in \mathcal{S}^{\otimes N}_k} C_{\pi,\sigma}  W _\pi \tr[]{W_\sigma O}. 
\label{eq:PhikN}
\end{align}
Here, $\pi=\bigotimes_{i=1}^N \pi_i \in \mathcal{S}^{\otimes N}_k $ and  $\sigma=\bigotimes_{i=1}^N \sigma_i \in \mathcal{S}^{\otimes N}_k $ are tensor products of permutations and the corresponding operators $W_\pi\equiv  \bigotimes W_{\pi_i}$ act locally on the $k$-fold copy space $\mathscr{h}^{\otimes k}$ of the individual qubits. The coefficients  $C_{\pi,\sigma} \equiv  \prod_{i=1}^N C_{\pi_i,\sigma_i}$  are determined by products of elements of the Weingarten matrix $C$.  
To proof Eq.~\eqref{eq:PhikN}, we first note that we can  expand an arbitrary operator $O$ in a product basis of the $N$ qudits $O=\sum_{\alpha} c_\alpha O_{\alpha_1} \otimes \dots \otimes O_{\alpha_N}$. By linearity, it suffices thus to restrict to $O$ being a tensor product of local operators, i.e.~$O=\bigotimes_{i=1}^N O_i$. We find
\begin{align}
\Phi^{(k)}_N \left(\bigotimes_{i=1}^N O_i \right) &= \bigotimes_{i=1}^N \overline{ \left( U_i^{\otimes k} \right) ^\dagger   O_i \; U_i^{\otimes k} } \nonumber  \\
&=  \bigotimes_{i=1}^N\Phi^{(k)}_1 (O_i)  \label{eq:PhiNProd} \\
&= \bigotimes_{i=1}^N \sum_{\pi_i,\sigma_i \in S_k} C_{\pi_i,\sigma_i}  W _{\pi_i} \tr[]{W_{\sigma_i} O_i}    \nonumber 
\end{align}
and from the last line,  Eq.~\eqref{eq:PhikN} follows. 

\begin{figure}
	\includegraphics[width=0.66\linewidth]{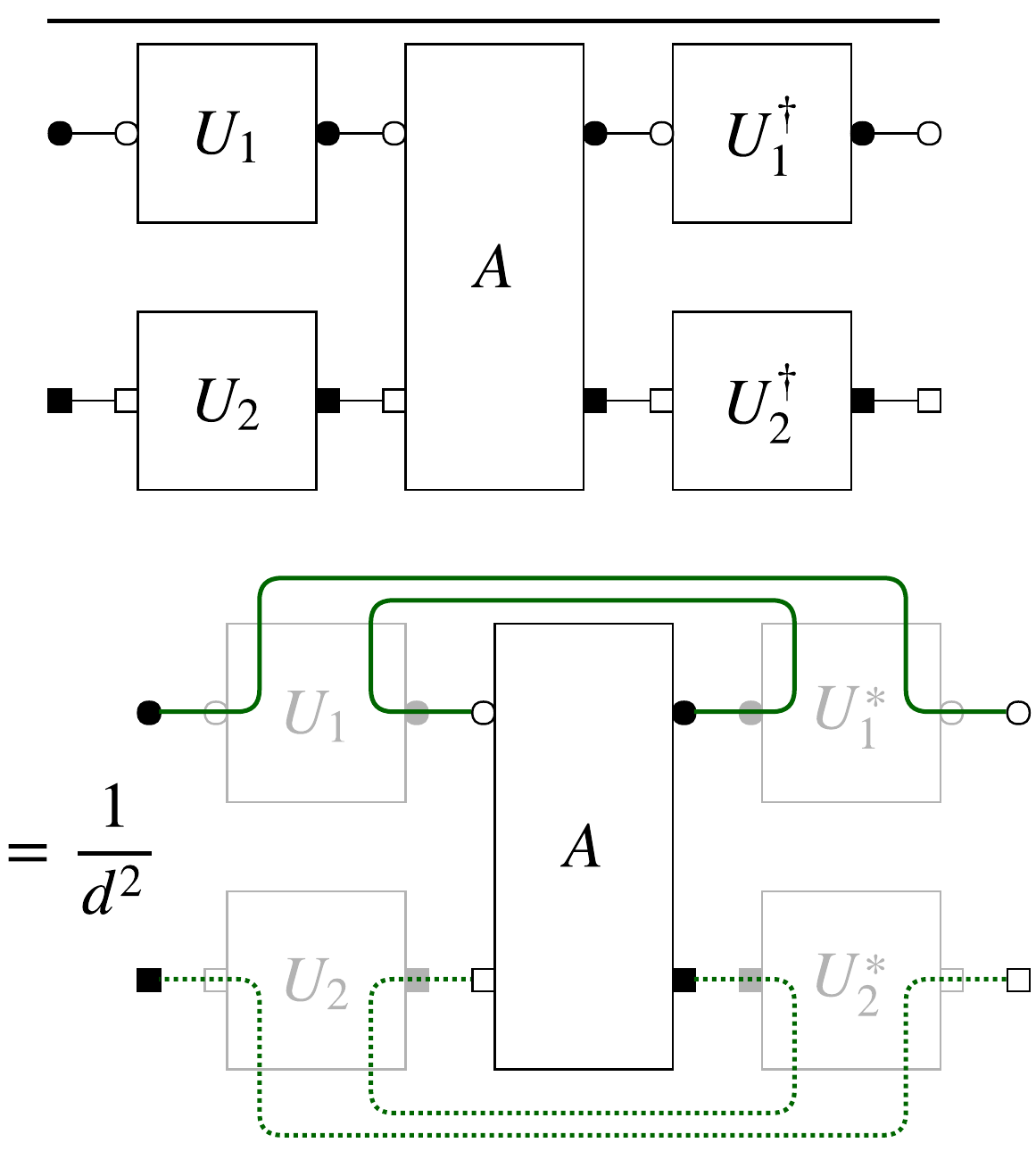}
	\caption{Graphical evaluation of $\Phi^{(1)}_2$. We evaluate graphically the matrix element $\bra{ij} \Phi^{(1)}_2(U) \ket{i'j'}$ where only decorations of the same type, corresponding to the same random unitary are connected. For clarity the boxes describing the basis states $\bra{ij}$, $\ket{i'j'}$ have  been omitted, only their decorations are kept. } 
		\label{fig:localunitaries}
\end{figure}

 { A graphical calculus to treat composite systems with multiple independent random unitaries has been introduced in Ref.~\cite{Vermersch2018a} to prove the central identities of a proposal for measuring out-of-time-ordered correlators with local random unitaries.}
To extend the graphical calculus presented here in Sec.~\ref{sec:diagram} one  introduces decorations of different types (circles, boxes, \dots) and connects only decorations of the same type.  For an example, we refer to Fig.~\ref{fig:localunitaries} where $\Phi^{(1)}_2$ is evaluated.

\section{Second order correlations - Purity and overlap of quantum states}
\label{sec:qst}

In this section, we apply our formalism to derive  Eqs.~\eqref{eq:AkkG} and \eqref{eq:Akkpur} allowing the estimation of the purity of arbitrary quantum states using the experimental protocol given in Sec.~\ref{sec:prot}. Furthermore, we discuss an  extension of the protocol given in Sec.~\ref{sec:prot}, to estimate the overlap of two distinct quantum states. Higher order functionals of the density matrix are discussed in the appendix \ref{sec:app}.

\subsection{Proof of the main result}
\label{sec:proof}

We now prove Eqs.~\eqref{eq:AkkG} and \eqref{eq:Akkpur}. We first note that the case (i) of global random unitaries  can be viewed as a single qu$\mathcal{D}$it with  dimension $\mathcal{D}_A$. Then, the first line of Eq.~\eqref{eq:AkkG} follows directly from Eq.~\eqref{eq:Akkpur}  by setting $N_A=1$ and $d\rightarrow \mathcal{D}_A=d^{N_A}$, and the second line by using $\sum_{\vec{s}_A,\vec{s}_A'\neq \vec{s}_A} P_U(\vec{s}'_A)=1-P_U(\vec{s}_A)$.
 Thus, we consider in the following the case (ii) of independent local random unitaries applied to a composite system of $N_A$ qudits with arbitrary local dimension $d$.  For simplicity of notation, we drop the subscript $A$.

We first note that the ensemble average of  second order cross-correlation of outcome probabilities of randomized measurements can be rewritten  as an expectation value of an operator $O$ acting on two ``virtual copies'' and the twirled state $\Phi_N^{(2)}( {\rho ^{\otimes 2} } )$ [with $\Phi_N^{(2)}$ defined in Eq.~\eqref{eq:PhikN}].
For arbitrary  coefficients
 $O_{\vec{s},\vec{s}'}$, it holds
\begin{align}
\sum_{\vec{s},\vec{s}'} &O_{\vec{s},\vec{s}'} \; \overline{  P_U(\vec{s}) P_U(\vec{s}') } \nonumber \\
&= \tr[]{\sum_{\vec{s},\vec{s}'} O_{\vec{s},\vec{s}'}  \ketbra{\vec{s}}{\vec{s}} \otimes  \ketbra{\vec{s}'}{\vec{s}'} \overline{ U^{\otimes 2}\rho \otimes \rho  \left(U^\dagger\right)^{\otimes 2}}} \nonumber \\
&=\tr[]{O \; \Phi_N^{(2)}(\rho \otimes \rho)} \nonumber \\
&=\tr[]{\Phi_N^{(2)}(O) \;  \rho \otimes \rho},
\label{eq:dualchannel}
\end{align}
where we defined the operator $O=\sum_{\vec{s},\vec{s}'} O_{\vec{s},\vec{s}'} \ketbra{\vec{s}}{\vec{s}} \otimes\ketbra{\vec{s}'}{\vec{s}'}  $ and used  the self-duality of the channel $\Phi^{(2)*}_N=\Phi^{(2)}_N$. 
Secondly, we observe that, for an arbitrary quantum state $\rho$, the purity can be rewritten as
\begin{align}
	\tr[]{\rho^2} = \tr[]{\mathbb{S} \rho \otimes \rho} ,
	\label{eq:swappur}
\end{align}
where   $ \mathbb{S}=\sum_{\vec{s},\vec{s}'} \ketbra{\vec{s}'}{\vec{s}} \otimes\ketbra{\vec{s}}{\vec{s}'} = W_{(2,1)^{\otimes N}}=W_{(2,1)}^{\otimes N}$ is the swap operator acting on two ``virtual'' copies $\mathcal{H}^{\otimes N} \otimes \mathcal{H}^{\otimes N} $ of the Hilbert space of $N$ qudits. 
Comparing Eqs. \eqref{eq:dualchannel} and \eqref{eq:swappur}, our goal is thus to find  coefficients $O_{\vec{s},\vec{s}'}$ of the operator $O$ such that
\begin{align}
\Phi_N^{(2)}(O) = \mathbb{S}  .
\end{align}
Since  $\Phi_N^{(2)}( o^{\otimes N}) =   \left(\Phi_1^{(2)}(o) \right)^{\otimes N}$ factorizes for an operator $O=\otimes_{i=1}^N o$ [Eq.~\eqref{eq:PhiNProd}], it is sufficient to find local operators $o= \sum_{s,s'=1}^d  o_{s,s'}\ketbra{s}{s} \otimes\ketbra{s'}{s'} $ which fulfill
\begin{align}
\Phi_1^{(2)}(o) = W_{(2,1)}   .
\end{align}
Using Eq.~\eqref{eq:Phik1},  this is equivalent to
\begin{align}
  \tr[]{W_\sigma o} & {=}  (C_{\sigma,(2,1)})^{-1} = d^{\sharp\text{cycles}(\sigma \cdot (2,1))}  \quad \forall \sigma \in \mathcal{S}_2 \; .
  \label{eq:S1}
\end{align}
Inserting the ansatz $o= \sum_{s,s'=1}^d  o_{s,s'}\ketbra{s}{s} \otimes\ketbra{s'}{s'} $ into Eq.~\eqref{eq:S1}, we find the following equations to be satisfied by the coefficients $o_{s,s'}$
\begin{align*}	
\text{and} \begin{split}\tr[]{W_{(1,2)} o}  &= \sum_{s,s'=1}^d o_{s,s'} \stackrel{!}{=} d  \nonumber \\
 \quad \tr[]{W_{(2,1)} o}  & = \sum_{s=1}^d o_{s,s} \stackrel{!}{=} d^2 .
 \end{split}
\end{align*}
These are satisfied by the simple choice 
\begin{align*}
o_{s,s'} =(d +1) \delta_{s,s'}  - 1 = d (-d)^{-D_G[s,s']} ,
\end{align*}
where $D_G[s,s']$ is the Hamming distance of the states $s$ and $s'$ of the single qudit, i.e. $D_G[s,s]=0$  and $D_G[s,s']=1$ if $s\neq s'$.
On the composite system, we then simply choose
\begin{align}
	O=o^{\otimes N} = d^N \sum_{\vec{s},\vec{s}'} (-d)^{-D[\vec{s},\vec{s}']}  \ketbra{\vec{s}}{\vec{s}} \otimes\ketbra{\vec{s}'}{\vec{s}'}  ,
	\label{eq:AkkN}
\end{align}
where $D[\vec{s},\vec{s}']=\sum_{i=1}^N D_G[s_i,s\label{key}'_i]$ is the Hamming distance of the states $\vec{s}$ and $\vec{s}'$ of $N$ qudits.
This leads directly to Eq. \eqref{eq:Akkpur}.

\subsection{Scaling of statistical errors}

In the previous sections, we have shown how to access the purity of an arbitrary quantum state $\rho_A$ from the ensemble average over cross correlations of outcome probabilities of randomized measurements. Here, we discuss the statistical errors arising in  an experiment due to a finite number $N_U$ of unitaries  to estimate the ensemble average and a finite number $N_M$ of measurements per random unitary to estimate the probabilities $P_U(\mathbf{s}_A)$. 

\begin{figure}
	\centering
	\includegraphics[width=0.99\linewidth]{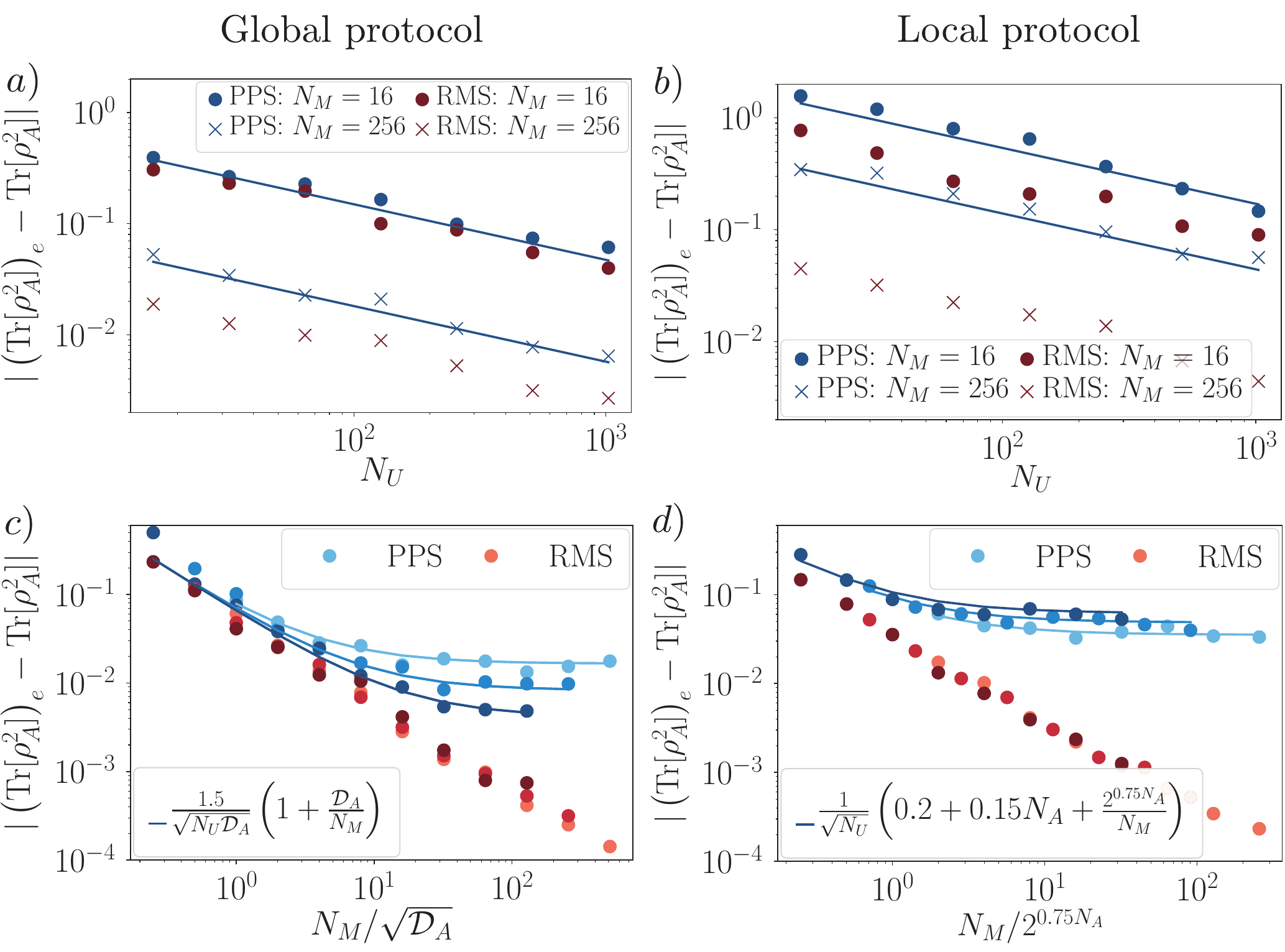}
	\caption{Statistical errors of the estimated purity of a pure product state (PPS) and a random mixed state (RMS). Panels a) and b) display the statistical errors of the estimated purity using a) global unitaries  and b) local unitaries in a system of $N_A=8$ qubits, as a function of $N_U$ for $N_M=16$ (dots) and $N_M=256$ (crosses). Hilbert space dimension is $\mathcal{D}=2^8$.  In panels c) (global unitaries) and d) (local unitaries) the number of unitaries $N_U=512$ is fixed, and the statistical error is shown as function of the number of measurements $N_M$. The Hilbert space dimension (number of qubits $N_A$) increases with darkness of the colors, $\mathcal{D}_A=2^4,2^6,2^8$. Solid lines are calculated from the given scaling laws. Random unitaries are sampled directly from CUE \cite{Mezzadri:2007}. The RMS of $N_A=4,6,8$ qubits has been obtained by applying a Haar random unitary to a pure product state consisting of $12$ qubits, and tracing out the residual $8,6,4$ degrees of freedom. }
	\label{fig:ErrorsPur}
\end{figure}

For the protocol utilizing global random unitaries, we found in Ref.~\cite{Vermersch:2018} analytically a scaling law of the typical statistical error of the estimated purity of a density matrix $\rho_A$ in a Hilbert space with dimension $\mathcal{D}_A$
\begin{align}
|\left( \tr[]{\rho_A^2} \right)_e - \tr[]{\rho_A^2} | \sim \frac{1}{\sqrt{N_U \mathcal{D}_A}} \left( c_1 + c_2 \frac{\mathcal{D}_A}{N_M} \right) .
\label{eq:globpursc}
\end{align}
where $c_1$ and $c_2$ are constants of $\mathcal{O}(1)$ which are largest for pure states. Thus the number of measurements per random unitary required to estimate the purity up to an error of $1/\sqrt{N_U}$ scales as $N_M \sim \sqrt{\mathcal{D}_A}$. The scaling behavior with $N_U$ is hereby a direct consequence of the central limit theorem, whereas the scaling with $N_M$ is directly related to probability of finding doublons when sampling with replacement from a finite probability distribution (the so-called birthday paradox \cite{Blinder2013}). Note, that we use unbiased estimators to infer the squared outcome probabilities $P_U(\mathbf{s}_A)^2$ from a finite number of measurements $N_M$ \cite{Vermersch:2018}. 
The analytical results are supported by numerical simulations presented in Fig.~\ref{fig:ErrorsPur}, panels a) and c) where the average statistical error, extracted from 100 numerical experiments, is shown, for a pure and a mixed state. The  lines are calculated from the scaling law \eqref{eq:globpursc}. Clearly, the statistical error of the mixed state is smaller, which is explained by the fact that fluctuations across the unitary ensemble are reduced for mixed states (with vanishing fluctuations for the maximally mixed state $\rho_{\text{max}}\sim \mathbb{1}/\mathcal{D}_A$).

For the protocol utilizing local random unitaries, we find, from numerical simulations, see Fig.~\ref{fig:ErrorsPur}, panels b) and d), for a pure product state of $N_A$ qubits, a scaling law
\begin{align}
|\left( \tr[]{\rho_A^2} \right)_e - \tr[]{\rho_A^2} | \sim \frac{1}{\sqrt{N_U }} \left( c_3+ \frac{2^{0.75 N_A}}{N_M} \right),
\end{align}
where $c_3=\mathcal{O}(N_A)$. The scaling with $N_U$ follows the behavior expected from the central limit theorem. The number of measurements to estimate the purity up to an error $\sim 1/\sqrt{N_U}$ scales as $N_M \sim 2^{0.75 N_A}$, and is thus larger than for the global protocol.  However,  in contrast to the global protocol, the statistical errors  for entangled states of $N_A$ qubits are reduced  which is explained by the fact that here the reduced density matrices for subsystems are mixed, and thus fluctuations across the unitary ensemble are locally reduced.
Similar as in the global protocol, this holds  true if   $\rho_{A}$ itself is mixed.

We note that on the one hand,   the local protocol (ii) is more prone to statistical errors compared to the global protocol (i). On the other hand, we can obtain  by restriction from the occupation probabilities $P_U(\mathbf{s}_{A})$ of basis states $\ket{\mathbf{s}_{A}}$ of the (sub-)system $A$, the occupation probabilities $P_{U}(\mathbf{s}_{A'})$ of basis  states $\ket{\mathbf{s}_{A'}}$ of an arbitrary subsystem ${A'} \subseteq A$. 
This is not possible for the global protocol, where the applied random unitary $U_{A}$ randomizing the entire Hilbert space $\mathcal{H}_A$ of $\rho_A$  predefines the (sub-) system $A$ of interest.

\subsection{Measurement of the overlap of quantum states}
 
In this subsection, we discuss a natural extension of the protocol presented in Sec.~\ref{sec:prot} to estimate the overlap $\tr[]{\rho_1\rho_2}$ of two distinct quantum states $\rho_1$ and $\rho_2$ of $N$ qudits in Hilbertspace $\mathcal{H}$ of dimension $\mathcal{D}=d^N$.  To obtain the overlap $\tr[]{\rho_1\rho_2}$, one applies the experimental sequence described in Sec.~\ref{sec:prot} twice with the same set of random unitaries, starting first with the state $\rho_1$ and secondly with the state $\rho_2$. Repeated for many random unitaries $U$, this provides the set of occupation probabilities $P^{(1)}_U(\vec{s})= \tr[]{U\rho_1 U^\dagger \ketbra{\vec{s}}{\vec{s}}}$ and 
$P^{(2)}_U(\vec{s})= \tr[]{U\rho_2 U^\dagger \ketbra{\vec{s}}{\vec{s}}}$ of the computational basis states $\ket{\vec{s}}$.  From cross correlations, the overlap is estimated. Generalizing Eqs.~\eqref{eq:AkkG} and \eqref{eq:Akkpur}, one finds, if (i) global random unitaries have been used
\begin{align}
\tr[]{\rho_1 \rho_2} =\mathcal{D}\sum_{\vec{s},\vec{s}'} (-\mathcal{D})^{-D_G[\vec{s},\vec{s}']}  \; \overline{  P^{(1)}_U(\vec{s}) P^{(2)}_U(\vec{s}') } 
\label{eq:ovg}
	\end{align}
	and (ii) for  local random unitaries
	\begin{align}
	\tr[]{\rho_1 \rho_2} = d^{N} \sum_{\vec{s},\vec{s}'} (-d)^{-D[\vec{s},\vec{s}']}  \; \overline{  P^{(1)}_U(\vec{s}) P^{(2)}_U(\vec{s}') } . 
	\label{eq:ovl}
	\end{align}
These equations follow directly from the proof presented in Sec.~\ref{sec:proof}, using that $\tr[]{\rho_1\rho_2} = \tr[]{\mathbb{S}\rho_1 \otimes \rho_2}$.

We note that this protocol enables a  measurement of the Loschmidt echo 
$ \left| \langle \psi_0 | e^{i H_2 t / \hbar} e^{-i H_1 t /	\hbar} | \psi_0 \rangle \right|^2 $  
\cite{Goussev2012} without the necessity of implementing time reversed operations or ancilla degrees of freedom; the protocol is outlined in the following: In a first experiment,  $\ket{\psi_0}$ is  evolved  forward in time with Hamiltonian $H_1$ and after the application of a random unitary $U$, the probabilities $P^{(1)}_U(\vec{s})=| \braket{ \vec{s}  |  U e^{-i H_1 t /	\hbar} |\psi_0}|^2  $  for the basis states $\ket{\mathbf{s}}$ are measured. This is then repeated with $H_2$, and \emph{same} random unitary $U$, to obtain  $P^{(2)}_U(\vec{s})=| \braket{ \vec{s}  |  U e^{-i H_2 t /	\hbar} |\psi_0}|^2  $.
The overlap $ \left| \langle \psi_0 | e^{i H_2 t / \hbar} e^{-i H_1 t /	\hbar} | \psi_0 \rangle \right|^2 $  is finally  inferred from cross-correlations over many random  unitaries, according to Eqs.~\eqref{eq:ovg} and \eqref{eq:ovl}.
Thus, there is no the necessity to implement the time reversed evolution operator $e^{i H_2 t / \hbar}$  in the experiment.

We further remark that this protocol can be used to check the stability of an experiment  against drifts, by measuring the overlap of two quantum states $\rho_1$ and $\rho_2$ which are prepared in the same way, but at different instances of time.

\section{Randomized quantum state tomography}
\label{sec:tom}

In this section, we describe a protocol to perform full quantum state tomography, based on statistical correlation of randomized measurements. For  global random unitaries,  this protocol was first described in Ref.~\cite{Ohliger2013} in the context of atomic Hubbard models. Here, we focus on spin models and extend the protocol to composite systems of many qudits where local unitaries are applied, and investigate in detail the scaling of the required number of measurements to estimate the density matrix up to a fixed statistical error with the Hilbert space dimension (the number of constituents).

We consider a quantum state $\rho_A$, which can be a reduced state $\rho_A=\tr[\mathcal{S} \backslash A]{\rho}$ of a subsystem $A\subseteq \mathcal{S}$, defined in the Hilbert  space $\mathcal{H}_A=\mathscr{h}^{\otimes N_A}$ with total dimension $\mathcal{D}_A=d^{N_A}$. 
The use of randomized measurements to perform quantum state tomography  is based on the  observation that 
\begin{align}
\rho_A = \tr[2]{\mathbb{S}  \;\id_{\mathcal{H}_A} \otimes \rho_A}  
\end{align}
where $S$ is the swap operator and the partial trace is taken over the second ``copy'. 
Using the results of Sec.~\ref{sec:proof}, this gives immediately rise to a measurement protocol to perform quantum state tomography using randomized measurements. Employing the ensemble average over \emph{global random} unitaries, randomizing the entire Hilbert space $\mathcal{H}_A$, we find
\begin{align}
\begin{split}
\rho_A &= \tr[2]{\Phi^{(2)}(O)  \;\id_{\mathcal{H}_A} \otimes \rho_A}  \\
& {=\mathcal{D}_A \sum_{\vec{s}_A,\vec{s}_A'} (-\mathcal{D}_A)^{-D_G[\vec{s}_A,\vec{s}_A']}    \overline{P_U(\vec{s}_A) \; U^\dagger_A \ketbra{\vec{s}_A'}{\vec{s}_A'} U_A} \; }\\
& {=(\mathcal{D}_A+1)\sum_{\vec{s}_A} \overline{P_U(\vec{s}_A) \; U_A^\dagger \ketbra{\vec{s}_A}{\vec{s}_A} U_A} -\id_{\mathcal{H}_A} ,}
\end{split}
\label{eq:random_tom_G}
\end{align}
where the expression in the last line was  first obtained in Ref.~\cite{Ohliger2013}.
If we consider instead  local random unitaries $U=\bigotimes_{i=1}^{N_A} U_i$ where the $U_i$ are sampled independently from $\text{CUE}(\mathscr{h})$, we find 
\begin{align}
\rho_A &=\tr[2]{\Phi^{(2)}_{N_A}(O)  \,\id_{d^{N_A}} \otimes \rho_A} \nonumber \\& {=d^{N_A} \sum_{\vec{s}_A,\vec{s}_A'} (-d)^{-D[\vec{s}_A,\vec{s}_A']}    \overline{P_U(\vec{s}_A) \; U^\dagger_A \ketbra{\vec{s}'_A}{\vec{s}_A'} U_A} }.
\label{eq:random_tom}
\end{align}
The experimental protocol reads thus as follows.  Given the quantum state $\rho_A$  a random unitary $U_A$, being either local or global, is applied. Subsequently, a measurement in the computational basis is performed. This is repeated with the same random unitary, to access the occupation probabilities $P_U(\vec{s}_A)$ of the computational basis states $\ket{s}_A$. Additionally,  the random unitary $U_A$ is stored as a matrix in the computational basis. Finally, the ensemble average is performed, as in Eqs.~\eqref{eq:random_tom_G} and \eqref{eq:random_tom}. 

\begin{figure}
	\centering
	\includegraphics[width=0.99\linewidth]{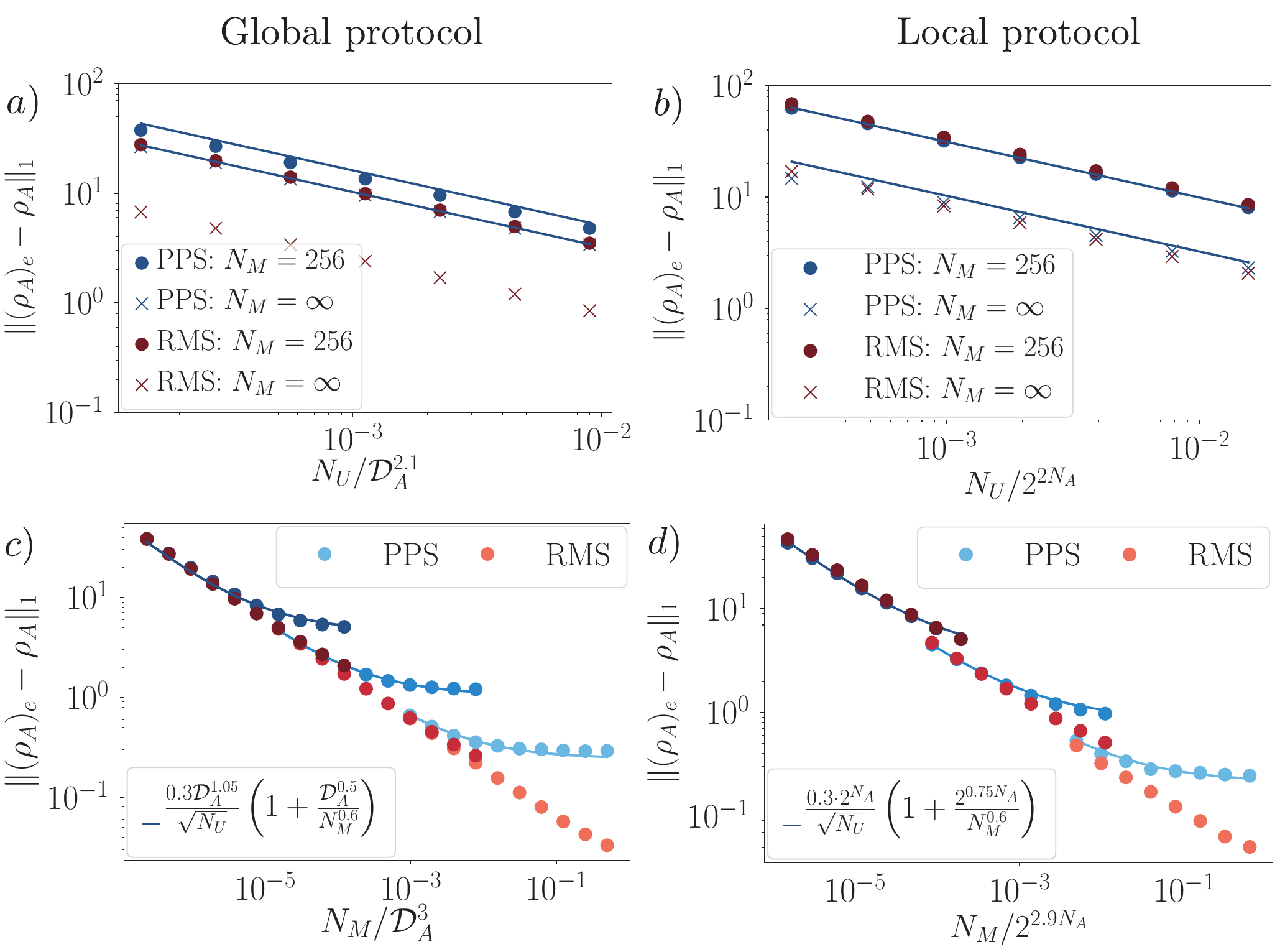}
	\caption{Statistical errors of the reconstructed density matrix for a pure product state (PPS) and a random mixed state (RMS). Panels a) and b) present the average deviation of the reconstructed density matrix  using a) global unitaries  and b) local unitaries in a system of $N_A=8$ qubits ($\mathcal{D}_A=2^8$), as a function of $N_U$ for $N_M = 256$ (dots) and $N_M = 
		\infty$, i.e., no projection noise, (crosses). In panels c) (global unitaries) and d) (local unitaries) the number of unitaries $N_U=512$ is fixed, and the average statistical error is shown as function of the number of measurements $N_M$. The Hilbert space dimension (number of qubits $N_A$) increases with darkness of the colors, $\mathcal{D}_A=2^4,2^6,2^8$. Solid lines are calculated from the given scaling laws. Random unitaries are sampled directly from CUE \cite{Mezzadri:2007}. The RMS of $N_A=4,6,8$ qubits has been obtained by applying a Haar random unitary to a pure product state consisting of $12$ qubits, and tracing out the residual $8,6,4$ degrees of freedom.}
	\label{fig:ErrorsTom}
\end{figure}

We emphasize that, in contrast to the previously discussed protocols, the tomographic reconstruction of $\rho_A$ from randomized measurements requires the explicit knowledge of the applied  unitaries in a specific basis, i.e.~not only their property of being sampled from an appropriate random matrix ensemble. The tomographic reconstruction is thus  prone to experimental imperfections which lead to a (random) mismatch between the ``applied'' random unitary and the one ``stored'' to  reconstruct $\rho_A$. In general, such decorrelation will appear as depolarizing noise, i.e.~induces a bias towards mixed state.  This is not the case for protocols which detect properties of $\rho_A$ which are invariant under unitary transformations, such as the purity estimation.

A crucial aspect for any measurement scheme providing tomographic reconstruction of a quantum state is the scaling of the required number of measurements with system size. In the protocol presented here, the accuracy of an estimation $(\rho_A)_e$, quantified by the trace distance $\|(\rho_A)_e-\rho\|_1$ to the true state $\rho_A$,  is determined by statistical errors originating from a finite number of random unitaries  $N_U$ and a finite number of measurements $N_M$ per random unitary.  
      In Fig.~\ref{fig:ErrorsTom}, the scaling behavior of the trace distance of the estimated density matrix $(\rho_A)_e$ for a pure product state $\rho_A$ is shown. For both protocols, we find that the numerical data, obtained with random unitaries sampled directly from the CUE \cite{Mezzadri:2007}, is well described by a scaling law
\begin{equation}
	\|(\rho_A)_e-\rho_A\|_1  \sim \frac{\mathcal{D}_A^a}{\sqrt{N_U}} \left( 1 + \frac{\mathcal{D}_A^b}{N_M^{0.6}} \right),
	\label{eq:errtom}
\end{equation}
where for the global protocol,  $a\approx 1.05$, $b\approx0.5$, and for the local protocol $a\approx 1.0$, $b\approx0.75$. From Eq.~\eqref{eq:errtom} it follows directly that, for pure product states,  the required number of random unitaries $N_U$  scales as $\mathcal{D}^2_A=2^{2N_A}$ and is thus comparable to the mininmal number of measurement settings in standard tomography \cite{Gross:2010}.  As shown in Fig.~\ref{fig:ErrorsTom}, the  statistical errors of an estimation of $\rho_A$ depend in general on the quantum state, where (absolute) errors smaller for mixed states (see also \cite{Brydges2018,Vermersch:2018}).  We note that it was shown in Ref.~\cite{Ohliger2013} that the above protocols for randomized state tomography can be combined with compressed sensing \cite{Gross:2010} to decrease the number of required measurements.

\section{Conclusion}

We have introduced statistical correlations of randomized measurements as a new tool to probe complex many-body quantum states. 
While we have focused on measurements protocol accessing bipartite entanglement of quantum states and their tomographic reconstructions, the underlying tools can be applied more generally. In a recent paper \cite{Vermersch2018a}, we use statistical correlations to design robust protocols  to measure out-of-time ordered correlation functions, without the necessity to implement time-reversed operations or ancilla degrees of freedom.  { A first experimental demonstration of this protocol was presented in Ref.~\cite{Nie2019} in a NMR quantum simulator consisting of four qubits.} In the future, the paradigm of statistical correlation could be extended to the measurement of order parameters in (symmetry protected) topological phases \cite{Wen2012,Pollmann2012,Haegeman2012,Shapourian2017} and the estimation of the entanglement spectrum \cite{Hui2008,Pichler2016,Dalmonte2018}. It could further be combined with techniques for randomized benchmarking of quantum computation \cite{Emerson2007,Knill2008,Wallman2015}.

Our protocols can be applied in state-of-the-art quantum simulators with single site read-out and control. As they rely on statistical correlations of many random measurements,  they are particularly suitable for systems with high repetition rates such as trapped-ions, superconducting qubits and Rydberg atoms. In atomic Hubbard models, one can take advantage of the possibility to prepare simultaneously many (independent) copies of the quantum systems to reduce the number of experimental runs.

Furthermore, it would be  interesting to extend the protocol based on local unitaries to models with (locally) conserved quantum numbers. For instance, local random unitaries with conserved particle number could be generated in atomic Hubbard models, by isolating pairs of sites and applying to each  a series of random quenches.

\begin{acknowledgements}We thank R. van Bijnen, R. Blatt, T. Brydges, C. Kokail, B. Kraus, L. Sieberer, and J. Yu for discussions. Within the European Union's Horizon 2020 research and innovation programme, this project has received funding under grant agreement No 817482 and from the European Research Council (ERC) under grant agreement No 741541. It is further supported by the ERC Synergy Grant UQUAM, the SFB FoQuS (FWF Project No. F4016-N23) and QTFLAG – QuantERA.  Numerical simulations were realized with QuTiP \cite{Johansson20131234}.
\end{acknowledgements}

%\bibliography{Bibliography}
\input{StatCorrRandomMeasurements.bbl}
\appendix

\section{Higher order R\'{e}nyi entropies}
\label{sec:app}

In this section, we discuss the estimation of $k$th-order functionals $\tr[]{\rho^k}$  ($k \in \mathbb{N}, k\geq 2$) from statistical correlations of globally randomized measurements, which are directly connected to $k$-th order R\'{e}nyi entropies $S^{(k)}(\rho) =1/(1-k)  \log_2\tr[]{\rho^k}$. 

The  experimental sequence is the same as described in section \ref{sec:prot}. Global random unitaries $U$ sampled from a \emph{unitary $k$-design} defined on the entire Hilbert space $\mathcal{H}$  are applied to the quantum state $\rho$ and subsequently occupation probabilities $P_U(\mathbf{s})$ are measured.  In Ref.~\cite{Vermersch:2018} we showed that the 
$k$-th moment $\overline{P_U(s)^k}$ is related to $\tr[]{\rho^k}$
\begin{align}
\overline{P_U(s)^k} =\frac{1}{\mathcal{D}_k}\sum_{ } C_{b_1,\dots, b_k} \prod_{l=1}^k \text{Tr}\left[\rho^{l} \right]^{b_l} 
\end{align}
where $\mathcal{D}_k =\prod_{i=0}^{k-1} ( \mathcal{D} +i)$ and   $C_{b_1,\dots, b_k}$  denotes the number of  permutations $\sigma \in S_k$ with $\textrm{typ} (\sigma )=1^{b_{1}}2^{b_{2}}\ldots k^{b_{k}}$  and is given by \cite{Bona2016} 
\begin{align}
C_{b_1,\dots, b_k} = {\frac  {k!}{b_{1}!\cdot b_{2}!\cdot \ldots \cdot b_{k}!\cdot 1^{{b_{1}}}\cdot 2^{{b_{2}}}\cdot \ldots \cdot k^{{b_{k}}}}} \; .
\end{align}
This result is easily recovered with the help of the results of section \ref{sec:rnduni}. We obtain 
\begin{align}
\overline{P_U(s)^k} &=\tr[]{\Phi_1^{(k)}\left(\rho^{\otimes k}\right) \ketbra{s}{s}^{\otimes k}}\nonumber \\
&=\tr[]{\Phi_1^{(k)}\left(\ketbra{s}{s}^{\otimes k} \right) \rho^{\otimes k} }\nonumber \\
&=\frac{1}{\mathcal{D}_k}\sum_{\sigma \in S_k} \tr[]{W_\sigma \rho^{\otimes k} } \\
&=\frac{1}{\mathcal{D}_k}\sum_{ } C_{b_1,\dots, b_k} \prod_{l=1}^k \text{Tr}\left[\rho^{l} \right]^{b_l} 
\label{eq:higher}
\end{align}
where we used that $\sum_{\pi \in S_k} C_{\pi \sigma }= 1/\prod_{i=0}^{k-1} ( \mathcal{D} +i) \equiv  1/ \mathcal{D}_k$ and $\tr[]{W_\sigma \ketbra{s}{s}^{\otimes k} }=1$ for any $\sigma \in S_k$. 
Since any unitary $k$-design is also a unitary $l$-design for $l<k$, we can reconstruct $\textrm{Tr}\left[\rho_A^{l} \right] $ for $l<k$ from lower order moments $P_U(s)^l$. Thus,  Eq.~\eqref{eq:higher} can be solved recursively to obtain $\tr[]{\rho^{k}}$.\\
In contrast to the case of global unitaries, the generalization of our method to access $\tr[]{\rho^k}$  with local unitaries is not straightforward, and will be studied in future work.

\end{document}

%% file: StatCorrRandomMeasurements.bbl

%% file: StatCorrRandomMeasurements.bbl
\begin{thebibliography}{63}%
\makeatletter
\providecommand \@ifxundefined [1]{%
 \@ifx{#1\undefined}
}%
\providecommand \@ifnum [1]{%
 \ifnum #1\expandafter \@firstoftwo
 \else \expandafter \@secondoftwo
 \fi
}%
\providecommand \@ifx [1]{%
 \ifx #1\expandafter \@firstoftwo
 \else \expandafter \@secondoftwo
 \fi
}%
\providecommand \natexlab [1]{#1}%
\providecommand \enquote  [1]{``#1''}%
\providecommand \bibnamefont  [1]{#1}%
\providecommand \bibfnamefont [1]{#1}%
\providecommand \citenamefont [1]{#1}%
\providecommand \href@noop [0]{\@secondoftwo}%
\providecommand \href [0]{\begingroup \@sanitize@url \@href}%
\providecommand \@href[1]{\@@startlink{#1}\@@href}%
\providecommand \@@href[1]{\endgroup#1\@@endlink}%
\providecommand \@sanitize@url [0]{\catcode `\\12\catcode `\$12\catcode
  `\&12\catcode `\#12\catcode `\^12\catcode `\_12\catcode `\%12\relax}%
\providecommand \@@startlink[1]{}%
\providecommand \@@endlink[0]{}%
\providecommand \url  [0]{\begingroup\@sanitize@url \@url }%
\providecommand \@url [1]{\endgroup\@href {#1}{\urlprefix }}%
\providecommand \urlprefix  [0]{URL }%
\providecommand \Eprint [0]{\href }%
\providecommand \doibase [0]{http://dx.doi.org/}%
\providecommand \selectlanguage [0]{\@gobble}%
\providecommand \bibinfo  [0]{\@secondoftwo}%
\providecommand \bibfield  [0]{\@secondoftwo}%
\providecommand \translation [1]{[#1]}%
\providecommand \BibitemOpen [0]{}%
\providecommand \bibitemStop [0]{}%
\providecommand \bibitemNoStop [0]{.\EOS\space}%
\providecommand \EOS [0]{\spacefactor3000\relax}%
\providecommand \BibitemShut  [1]{\csname bibitem#1\endcsname}%
\let\auto@bib@innerbib\@empty
%</preamble>
\bibitem [{\citenamefont {Preskill}(2018)}]{Preskill:2018}%
  \BibitemOpen
  \bibfield  {author} {\bibinfo {author} {\bibfnamefont {J.}~\bibnamefont
  {Preskill}},\ }\href {\doibase 10.22331/q-2018-08-06-79} {\bibfield
  {journal} {\bibinfo  {journal} {{Quantum}}\ }\textbf {\bibinfo {volume}
  {2}},\ \bibinfo {pages} {79} (\bibinfo {year} {2018})}\BibitemShut {NoStop}%
\bibitem [{\citenamefont {Bloch}\ \emph {et~al.}(2012)\citenamefont {Bloch},
  \citenamefont {Dalibard},\ and\ \citenamefont {Nascimb\`{e}ne}}]{Bloch:2012}%
  \BibitemOpen
  \bibfield  {author} {\bibinfo {author} {\bibfnamefont {I.}~\bibnamefont
  {Bloch}}, \bibinfo {author} {\bibfnamefont {J.}~\bibnamefont {Dalibard}}, \
  and\ \bibinfo {author} {\bibfnamefont {S.}~\bibnamefont {Nascimb\`{e}ne}},\
  }\href {https://www.nature.com/articles/nphys138} {\bibfield  {journal}
  {\bibinfo  {journal} {Nat.~Phys.}\ }\textbf {\bibinfo {volume} {8}},\
  \bibinfo {pages} {267} (\bibinfo {year} {2012})}\BibitemShut {NoStop}%
\bibitem [{\citenamefont {Blatt}\ and\ \citenamefont {Roos}(2012)}]{Blatt2012}%
  \BibitemOpen
  \bibfield  {author} {\bibinfo {author} {\bibfnamefont {R.}~\bibnamefont
  {Blatt}}\ and\ \bibinfo {author} {\bibfnamefont {C.~F.}\ \bibnamefont
  {Roos}},\ }\href {https://www.nature.com/articles/nphys2252} {\bibfield
  {journal} {\bibinfo  {journal} {Nature Physics}\ }\textbf {\bibinfo {volume}
  {8}},\ \bibinfo {pages} {277} (\bibinfo {year} {2012})}\BibitemShut {NoStop}%
\bibitem [{\citenamefont {Browaeys}\ \emph {et~al.}(2016)\citenamefont
  {Browaeys}, \citenamefont {Barredo},\ and\ \citenamefont
  {Lahaye}}]{Browaeys:2016}%
  \BibitemOpen
  \bibfield  {author} {\bibinfo {author} {\bibfnamefont {A.}~\bibnamefont
  {Browaeys}}, \bibinfo {author} {\bibfnamefont {D.}~\bibnamefont {Barredo}}, \
  and\ \bibinfo {author} {\bibfnamefont {T.}~\bibnamefont {Lahaye}},\ }\href
  {http://stacks.iop.org/0953-4075/49/i=15/a=152001} {\bibfield  {journal}
  {\bibinfo  {journal} {Journal of Physics B: Atomic, Molecular and Optical
  Physics}\ }\textbf {\bibinfo {volume} {49}},\ \bibinfo {pages} {152001}
  (\bibinfo {year} {2016})}\BibitemShut {NoStop}%
\bibitem [{\citenamefont {Gambetta}\ \emph {et~al.}(2017)\citenamefont
  {Gambetta}, \citenamefont {Chow},\ and\ \citenamefont
  {Steffen}}]{Gambetta:2017}%
  \BibitemOpen
  \bibfield  {author} {\bibinfo {author} {\bibfnamefont {J.~M.}\ \bibnamefont
  {Gambetta}}, \bibinfo {author} {\bibfnamefont {J.~M.}\ \bibnamefont {Chow}},
  \ and\ \bibinfo {author} {\bibfnamefont {M.}~\bibnamefont {Steffen}},\ }\href
  {https://www.nature.com/articles/s41534-016-0004-0} {\bibfield  {journal}
  {\bibinfo  {journal} {npj Quantum Information}\ }\textbf {\bibinfo {volume}
  {3}},\ \bibinfo {pages} {2} (\bibinfo {year} {2017})}\BibitemShut {NoStop}%
\bibitem [{\citenamefont {Horodecki}\ \emph {et~al.}(2009)\citenamefont
  {Horodecki}, \citenamefont {Horodecki}, \citenamefont {Horodecki},\ and\
  \citenamefont {Horodecki}}]{Horodecki:2009}%
  \BibitemOpen
  \bibfield  {author} {\bibinfo {author} {\bibfnamefont {R.}~\bibnamefont
  {Horodecki}}, \bibinfo {author} {\bibfnamefont {P.}~\bibnamefont
  {Horodecki}}, \bibinfo {author} {\bibfnamefont {M.}~\bibnamefont
  {Horodecki}}, \ and\ \bibinfo {author} {\bibfnamefont {K.}~\bibnamefont
  {Horodecki}},\ }\href {\doibase 10.1103/RevModPhys.81.865} {\bibfield
  {journal} {\bibinfo  {journal} {Rev. Mod. Phys.}\ }\textbf {\bibinfo {volume}
  {81}},\ \bibinfo {pages} {865} (\bibinfo {year} {2009})}\BibitemShut
  {NoStop}%
\bibitem [{\citenamefont {{H{\"a}ffner}}\ \emph {et~al.}(2005)\citenamefont
  {{H{\"a}ffner}}, \citenamefont {{H{\"a}nsel}}, \citenamefont {{Roos}},
  \citenamefont {{Benhelm}}, \citenamefont {{Chek-Al-Kar}}, \citenamefont
  {{Chwalla}}, \citenamefont {{K{\"o}rber}}, \citenamefont {{Rapol}},
  \citenamefont {{Riebe}}, \citenamefont {{Schmidt}}, \citenamefont {{Becher}},
  \citenamefont {{G{\"u}hne}}, \citenamefont {{D{\"u}r}},\ and\ \citenamefont
  {{Blatt}}}]{Haeffner2005}%
  \BibitemOpen
  \bibfield  {author} {\bibinfo {author} {\bibfnamefont {H.}~\bibnamefont
  {{H{\"a}ffner}}}, \bibinfo {author} {\bibfnamefont {W.}~\bibnamefont
  {{H{\"a}nsel}}}, \bibinfo {author} {\bibfnamefont {C.~F.}\ \bibnamefont
  {{Roos}}}, \bibinfo {author} {\bibfnamefont {J.}~\bibnamefont {{Benhelm}}},
  \bibinfo {author} {\bibfnamefont {D.}~\bibnamefont {{Chek-Al-Kar}}}, \bibinfo
  {author} {\bibfnamefont {M.}~\bibnamefont {{Chwalla}}}, \bibinfo {author}
  {\bibfnamefont {T.}~\bibnamefont {{K{\"o}rber}}}, \bibinfo {author}
  {\bibfnamefont {U.~D.}\ \bibnamefont {{Rapol}}}, \bibinfo {author}
  {\bibfnamefont {M.}~\bibnamefont {{Riebe}}}, \bibinfo {author} {\bibfnamefont
  {P.~O.}\ \bibnamefont {{Schmidt}}}, \bibinfo {author} {\bibfnamefont
  {C.}~\bibnamefont {{Becher}}}, \bibinfo {author} {\bibfnamefont
  {O.}~\bibnamefont {{G{\"u}hne}}}, \bibinfo {author} {\bibfnamefont
  {W.}~\bibnamefont {{D{\"u}r}}}, \ and\ \bibinfo {author} {\bibfnamefont
  {R.}~\bibnamefont {{Blatt}}},\ }\href {\doibase 10.1038/nature04279}
  {\bibfield  {journal} {\bibinfo  {journal} {\nat}\ }\textbf {\bibinfo
  {volume} {438}},\ \bibinfo {pages} {643} (\bibinfo {year}
  {2005})}\BibitemShut {NoStop}%
\bibitem [{\citenamefont {Gross}\ \emph {et~al.}(2010)\citenamefont {Gross},
  \citenamefont {Liu}, \citenamefont {Flammia}, \citenamefont {Becker},\ and\
  \citenamefont {Eisert}}]{Gross:2010}%
  \BibitemOpen
  \bibfield  {author} {\bibinfo {author} {\bibfnamefont {D.}~\bibnamefont
  {Gross}}, \bibinfo {author} {\bibfnamefont {Y.-K.}\ \bibnamefont {Liu}},
  \bibinfo {author} {\bibfnamefont {S.~T.}\ \bibnamefont {Flammia}}, \bibinfo
  {author} {\bibfnamefont {S.}~\bibnamefont {Becker}}, \ and\ \bibinfo {author}
  {\bibfnamefont {J.}~\bibnamefont {Eisert}},\ }\href {\doibase
  10.1103/PhysRevLett.105.150401} {\bibfield  {journal} {\bibinfo  {journal}
  {Phys. Rev. Lett.}\ }\textbf {\bibinfo {volume} {105}},\ \bibinfo {pages}
  {150401} (\bibinfo {year} {2010})}\BibitemShut {NoStop}%
\bibitem [{\citenamefont {Lanyon}\ \emph {et~al.}(2017)\citenamefont {Lanyon},
  \citenamefont {Maier}, \citenamefont {Holz\"apfel}, \citenamefont
  {Baumgratz}, \citenamefont {Hempel}, \citenamefont {Jurcevic}, \citenamefont
  {Dhand}, \citenamefont {Buyskikh}, \citenamefont {Daley}, \citenamefont
  {Cramer}, \citenamefont {Plenio}, \citenamefont {Blatt},\ and\ \citenamefont
  {Roos}}]{Lanyon:2017}%
  \BibitemOpen
  \bibfield  {author} {\bibinfo {author} {\bibfnamefont {B.~P.}\ \bibnamefont
  {Lanyon}}, \bibinfo {author} {\bibfnamefont {C.}~\bibnamefont {Maier}},
  \bibinfo {author} {\bibfnamefont {M.}~\bibnamefont {Holz\"apfel}}, \bibinfo
  {author} {\bibfnamefont {T.}~\bibnamefont {Baumgratz}}, \bibinfo {author}
  {\bibfnamefont {C.}~\bibnamefont {Hempel}}, \bibinfo {author} {\bibfnamefont
  {P.}~\bibnamefont {Jurcevic}}, \bibinfo {author} {\bibfnamefont
  {I.}~\bibnamefont {Dhand}}, \bibinfo {author} {\bibfnamefont {A.~S.}\
  \bibnamefont {Buyskikh}}, \bibinfo {author} {\bibfnamefont {A.~J.}\
  \bibnamefont {Daley}}, \bibinfo {author} {\bibfnamefont {M.}~\bibnamefont
  {Cramer}}, \bibinfo {author} {\bibfnamefont {M.~C.}\ \bibnamefont {Plenio}},
  \bibinfo {author} {\bibfnamefont {R.}~\bibnamefont {Blatt}}, \ and\ \bibinfo
  {author} {\bibfnamefont {C.~F.}\ \bibnamefont {Roos}},\ }\href
  {https://www.nature.com/articles/nphys4244} {\bibfield  {journal} {\bibinfo
  {journal} {Nat. Phys.}\ }\textbf {\bibinfo {volume} {13}},\ \bibinfo {pages}
  {1158} (\bibinfo {year} {2017})}\BibitemShut {NoStop}%
\bibitem [{\citenamefont {Torlai}\ \emph {et~al.}(2018)\citenamefont {Torlai},
  \citenamefont {Mazzola}, \citenamefont {Carrasquilla}, \citenamefont
  {Troyer}, \citenamefont {Melko},\ and\ \citenamefont {Carleo}}]{Torlai:2018}%
  \BibitemOpen
  \bibfield  {author} {\bibinfo {author} {\bibfnamefont {G.}~\bibnamefont
  {Torlai}}, \bibinfo {author} {\bibfnamefont {G.}~\bibnamefont {Mazzola}},
  \bibinfo {author} {\bibfnamefont {J.}~\bibnamefont {Carrasquilla}}, \bibinfo
  {author} {\bibfnamefont {M.}~\bibnamefont {Troyer}}, \bibinfo {author}
  {\bibfnamefont {R.}~\bibnamefont {Melko}}, \ and\ \bibinfo {author}
  {\bibfnamefont {G.}~\bibnamefont {Carleo}},\ }\href {\doibase
  10.1038/s41567-018-0048-5} {\bibfield  {journal} {\bibinfo  {journal} {Nat.
  Phys.}\ }\textbf {\bibinfo {volume} {14}},\ \bibinfo {pages} {447} (\bibinfo
  {year} {2018})}\BibitemShut {NoStop}%
\bibitem [{\citenamefont {Islam}\ \emph {et~al.}(2015)\citenamefont {Islam},
  \citenamefont {Ma}, \citenamefont {Preiss}, \citenamefont {Tai},
  \citenamefont {Lukin}, \citenamefont {Rispoli},\ and\ \citenamefont
  {Greiner}}]{Islam:2015}%
  \BibitemOpen
  \bibfield  {author} {\bibinfo {author} {\bibfnamefont {R.}~\bibnamefont
  {Islam}}, \bibinfo {author} {\bibfnamefont {R.}~\bibnamefont {Ma}}, \bibinfo
  {author} {\bibfnamefont {P.~M.}\ \bibnamefont {Preiss}}, \bibinfo {author}
  {\bibfnamefont {M.~E.}\ \bibnamefont {Tai}}, \bibinfo {author} {\bibfnamefont
  {A.}~\bibnamefont {Lukin}}, \bibinfo {author} {\bibfnamefont
  {M.}~\bibnamefont {Rispoli}}, \ and\ \bibinfo {author} {\bibfnamefont
  {M.}~\bibnamefont {Greiner}},\ }\href
  {https://www.nature.com/articles/nature15750} {\bibfield  {journal} {\bibinfo
   {journal} {Nature}\ }\textbf {\bibinfo {volume} {528}},\ \bibinfo {pages}
  {77} (\bibinfo {year} {2015})}\BibitemShut {NoStop}%
\bibitem [{\citenamefont {Kaufman}\ \emph {et~al.}(2016)\citenamefont
  {Kaufman}, \citenamefont {Tai}, \citenamefont {Lukin}, \citenamefont
  {Rispoli}, \citenamefont {Schittko}, \citenamefont {Preiss},\ and\
  \citenamefont {Greiner}}]{Kaufman:2016}%
  \BibitemOpen
  \bibfield  {author} {\bibinfo {author} {\bibfnamefont {A.~M.}\ \bibnamefont
  {Kaufman}}, \bibinfo {author} {\bibfnamefont {M.~E.}\ \bibnamefont {Tai}},
  \bibinfo {author} {\bibfnamefont {A.}~\bibnamefont {Lukin}}, \bibinfo
  {author} {\bibfnamefont {M.}~\bibnamefont {Rispoli}}, \bibinfo {author}
  {\bibfnamefont {R.}~\bibnamefont {Schittko}}, \bibinfo {author}
  {\bibfnamefont {P.~M.}\ \bibnamefont {Preiss}}, \ and\ \bibinfo {author}
  {\bibfnamefont {M.}~\bibnamefont {Greiner}},\ }\href {\doibase
  10.1126/science.aaf6725} {\bibfield  {journal} {\bibinfo  {journal}
  {Science}\ }\textbf {\bibinfo {volume} {353}},\ \bibinfo {pages} {794}
  (\bibinfo {year} {2016})}\BibitemShut {NoStop}%
\bibitem [{\citenamefont {Bovino}\ \emph {et~al.}(2005)\citenamefont {Bovino},
  \citenamefont {Castagnoli}, \citenamefont {Ekert}, \citenamefont {Horodecki},
  \citenamefont {Alves},\ and\ \citenamefont {Sergienko}}]{Ekert:2002}%
  \BibitemOpen
  \bibfield  {author} {\bibinfo {author} {\bibfnamefont {F.~A.}\ \bibnamefont
  {Bovino}}, \bibinfo {author} {\bibfnamefont {G.}~\bibnamefont {Castagnoli}},
  \bibinfo {author} {\bibfnamefont {A.}~\bibnamefont {Ekert}}, \bibinfo
  {author} {\bibfnamefont {P.}~\bibnamefont {Horodecki}}, \bibinfo {author}
  {\bibfnamefont {C.~M.}\ \bibnamefont {Alves}}, \ and\ \bibinfo {author}
  {\bibfnamefont {A.~V.}\ \bibnamefont {Sergienko}},\ }\href {\doibase
  10.1103/PhysRevLett.95.240407} {\bibfield  {journal} {\bibinfo  {journal}
  {Phys. Rev. Lett.}\ }\textbf {\bibinfo {volume} {95}},\ \bibinfo {pages}
  {240407} (\bibinfo {year} {2005})}\BibitemShut {NoStop}%
\bibitem [{\citenamefont {Daley}\ \emph {et~al.}(2012)\citenamefont {Daley},
  \citenamefont {Pichler}, \citenamefont {Schachenmayer},\ and\ \citenamefont
  {Zoller}}]{Daley2012}%
  \BibitemOpen
  \bibfield  {author} {\bibinfo {author} {\bibfnamefont {A.~J.}\ \bibnamefont
  {Daley}}, \bibinfo {author} {\bibfnamefont {H.}~\bibnamefont {Pichler}},
  \bibinfo {author} {\bibfnamefont {J.}~\bibnamefont {Schachenmayer}}, \ and\
  \bibinfo {author} {\bibfnamefont {P.}~\bibnamefont {Zoller}},\ }\href
  {https://journals.aps.org/prl/abstract/10.1103/PhysRevLett.109.020505}
  {\bibfield  {journal} {\bibinfo  {journal} {Phys. Rev. Lett.}\ }\textbf
  {\bibinfo {volume} {109}} (\bibinfo {year} {2012})}\BibitemShut {NoStop}%
\bibitem [{\citenamefont {Pichler}\ \emph {et~al.}(2013)\citenamefont
  {Pichler}, \citenamefont {Bonnes}, \citenamefont {Daley}, \citenamefont
  {Läuchli},\ and\ \citenamefont {Zoller}}]{Pichler:2013}%
  \BibitemOpen
  \bibfield  {author} {\bibinfo {author} {\bibfnamefont {H.}~\bibnamefont
  {Pichler}}, \bibinfo {author} {\bibfnamefont {L.}~\bibnamefont {Bonnes}},
  \bibinfo {author} {\bibfnamefont {A.~J.}\ \bibnamefont {Daley}}, \bibinfo
  {author} {\bibfnamefont {A.~M.}\ \bibnamefont {Läuchli}}, \ and\ \bibinfo
  {author} {\bibfnamefont {P.}~\bibnamefont {Zoller}},\ }\href
  {http://stacks.iop.org/1367-2630/15/i=6/a=063003} {\bibfield  {journal}
  {\bibinfo  {journal} {New Journal of Physics}\ }\textbf {\bibinfo {volume}
  {15}},\ \bibinfo {pages} {063003} (\bibinfo {year} {2013})}\BibitemShut
  {NoStop}%
\bibitem [{\citenamefont {Brydges}\ \emph {et~al.}(2019)\citenamefont
  {Brydges}, \citenamefont {Elben}, \citenamefont {Jurcevic}, \citenamefont
  {Vermersch}, \citenamefont {Maier}, \citenamefont {Lanyon}, \citenamefont
  {Zoller}, \citenamefont {Blatt},\ and\ \citenamefont {Roos}}]{Brydges2018}%
  \BibitemOpen
  \bibfield  {author} {\bibinfo {author} {\bibfnamefont {T.}~\bibnamefont
  {Brydges}}, \bibinfo {author} {\bibfnamefont {A.}~\bibnamefont {Elben}},
  \bibinfo {author} {\bibfnamefont {P.}~\bibnamefont {Jurcevic}}, \bibinfo
  {author} {\bibfnamefont {B.}~\bibnamefont {Vermersch}}, \bibinfo {author}
  {\bibfnamefont {C.}~\bibnamefont {Maier}}, \bibinfo {author} {\bibfnamefont
  {B.~P.}\ \bibnamefont {Lanyon}}, \bibinfo {author} {\bibfnamefont
  {P.}~\bibnamefont {Zoller}}, \bibinfo {author} {\bibfnamefont
  {R.}~\bibnamefont {Blatt}}, \ and\ \bibinfo {author} {\bibfnamefont {C.~F.}\
  \bibnamefont {Roos}},\ }\href {\doibase 10.1126/science.aau4963} {\bibfield
  {journal} {\bibinfo  {journal} {Science}\ }\textbf {\bibinfo {volume}
  {364}},\ \bibinfo {pages} {260} (\bibinfo {year} {2019})}\BibitemShut
  {NoStop}%
\bibitem [{\citenamefont {Elben}\ \emph {et~al.}(2018)\citenamefont {Elben},
  \citenamefont {Vermersch}, \citenamefont {Dalmonte}, \citenamefont {Cirac},\
  and\ \citenamefont {Zoller}}]{Elben:2018}%
  \BibitemOpen
  \bibfield  {author} {\bibinfo {author} {\bibfnamefont {A.}~\bibnamefont
  {Elben}}, \bibinfo {author} {\bibfnamefont {B.}~\bibnamefont {Vermersch}},
  \bibinfo {author} {\bibfnamefont {M.}~\bibnamefont {Dalmonte}}, \bibinfo
  {author} {\bibfnamefont {J.~I.}\ \bibnamefont {Cirac}}, \ and\ \bibinfo
  {author} {\bibfnamefont {P.}~\bibnamefont {Zoller}},\ }\href {\doibase
  10.1103/PhysRevLett.120.050406} {\bibfield  {journal} {\bibinfo  {journal}
  {Phys. Rev. Lett.}\ }\textbf {\bibinfo {volume} {120}},\ \bibinfo {pages}
  {050406} (\bibinfo {year} {2018})}\BibitemShut {NoStop}%
\bibitem [{\citenamefont {Zhang}\ \emph {et~al.}(2017)\citenamefont {Zhang},
  \citenamefont {Pagano}, \citenamefont {Hess}, \citenamefont {Kyprianidis},
  \citenamefont {Becker}, \citenamefont {Kaplan}, \citenamefont {Gorshkov},
  \citenamefont {Gong},\ and\ \citenamefont {Monroe}}]{Zhang:2017a}%
  \BibitemOpen
  \bibfield  {author} {\bibinfo {author} {\bibfnamefont {J.}~\bibnamefont
  {Zhang}}, \bibinfo {author} {\bibfnamefont {G.}~\bibnamefont {Pagano}},
  \bibinfo {author} {\bibfnamefont {P.~W.}\ \bibnamefont {Hess}}, \bibinfo
  {author} {\bibfnamefont {A.}~\bibnamefont {Kyprianidis}}, \bibinfo {author}
  {\bibfnamefont {P.}~\bibnamefont {Becker}}, \bibinfo {author} {\bibfnamefont
  {H.}~\bibnamefont {Kaplan}}, \bibinfo {author} {\bibfnamefont {A.~V.}\
  \bibnamefont {Gorshkov}}, \bibinfo {author} {\bibfnamefont {Z.-X.}\
  \bibnamefont {Gong}}, \ and\ \bibinfo {author} {\bibfnamefont
  {C.}~\bibnamefont {Monroe}},\ }\href
  {https://www.nature.com/articles/nature24654} {\bibfield  {journal} {\bibinfo
   {journal} {Nature}\ }\textbf {\bibinfo {volume} {551}},\ \bibinfo {pages}
  {601} (\bibinfo {year} {2017})}\BibitemShut {NoStop}%
\bibitem [{\citenamefont {Zeiher}\ \emph {et~al.}(2017)\citenamefont {Zeiher},
  \citenamefont {Choi}, \citenamefont {Rubio-Abadal}, \citenamefont {Pohl},
  \citenamefont {van Bijnen}, \citenamefont {Bloch},\ and\ \citenamefont
  {Gross}}]{Zeiher2017}%
  \BibitemOpen
  \bibfield  {author} {\bibinfo {author} {\bibfnamefont {J.}~\bibnamefont
  {Zeiher}}, \bibinfo {author} {\bibfnamefont {J.-y.}\ \bibnamefont {Choi}},
  \bibinfo {author} {\bibfnamefont {A.}~\bibnamefont {Rubio-Abadal}}, \bibinfo
  {author} {\bibfnamefont {T.}~\bibnamefont {Pohl}}, \bibinfo {author}
  {\bibfnamefont {R.}~\bibnamefont {van Bijnen}}, \bibinfo {author}
  {\bibfnamefont {I.}~\bibnamefont {Bloch}}, \ and\ \bibinfo {author}
  {\bibfnamefont {C.}~\bibnamefont {Gross}},\ }\href
  {https://journals.aps.org/prx/abstract/10.1103/PhysRevX.7.041063} {\bibfield
  {journal} {\bibinfo  {journal} {Physical Review X}\ }\textbf {\bibinfo
  {volume} {7}},\ \bibinfo {pages} {041063} (\bibinfo {year}
  {2017})}\BibitemShut {NoStop}%
\bibitem [{\citenamefont {Barredo}\ \emph {et~al.}(2018)\citenamefont
  {Barredo}, \citenamefont {Lienhard}, \citenamefont {De~Leseleuc},
  \citenamefont {Lahaye},\ and\ \citenamefont {Browaeys}}]{Barredo2018}%
  \BibitemOpen
  \bibfield  {author} {\bibinfo {author} {\bibfnamefont {D.}~\bibnamefont
  {Barredo}}, \bibinfo {author} {\bibfnamefont {V.}~\bibnamefont {Lienhard}},
  \bibinfo {author} {\bibfnamefont {S.}~\bibnamefont {De~Leseleuc}}, \bibinfo
  {author} {\bibfnamefont {T.}~\bibnamefont {Lahaye}}, \ and\ \bibinfo {author}
  {\bibfnamefont {A.}~\bibnamefont {Browaeys}},\ }\href
  {https://www.nature.com/articles/s41586-018-0450-2} {\bibfield  {journal}
  {\bibinfo  {journal} {Nature}\ }\textbf {\bibinfo {volume} {561}},\ \bibinfo
  {pages} {79} (\bibinfo {year} {2018})}\BibitemShut {NoStop}%
\bibitem [{\citenamefont {Guardado-Sanchez}\ \emph {et~al.}(2018)\citenamefont
  {Guardado-Sanchez}, \citenamefont {Brown}, \citenamefont {Mitra},
  \citenamefont {Devakul}, \citenamefont {Huse}, \citenamefont {Schau{\ss}},\
  and\ \citenamefont {Bakr}}]{Guardado2018}%
  \BibitemOpen
  \bibfield  {author} {\bibinfo {author} {\bibfnamefont {E.}~\bibnamefont
  {Guardado-Sanchez}}, \bibinfo {author} {\bibfnamefont {P.~T.}\ \bibnamefont
  {Brown}}, \bibinfo {author} {\bibfnamefont {D.}~\bibnamefont {Mitra}},
  \bibinfo {author} {\bibfnamefont {T.}~\bibnamefont {Devakul}}, \bibinfo
  {author} {\bibfnamefont {D.~A.}\ \bibnamefont {Huse}}, \bibinfo {author}
  {\bibfnamefont {P.}~\bibnamefont {Schau{\ss}}}, \ and\ \bibinfo {author}
  {\bibfnamefont {W.~S.}\ \bibnamefont {Bakr}},\ }\href
  {https://journals.aps.org/prx/abstract/10.1103/PhysRevX.8.021069} {\bibfield
  {journal} {\bibinfo  {journal} {Physical Review X}\ }\textbf {\bibinfo
  {volume} {8}},\ \bibinfo {pages} {021069} (\bibinfo {year}
  {2018})}\BibitemShut {NoStop}%
\bibitem [{\citenamefont {Keesling}\ \emph {et~al.}(2019)\citenamefont
  {Keesling}, \citenamefont {Omran}, \citenamefont {Levine}, \citenamefont
  {Bernien}, \citenamefont {Pichler}, \citenamefont {Choi}, \citenamefont
  {Samajdar}, \citenamefont {Schwartz}, \citenamefont {Silvi}, \citenamefont
  {Sachdev} \emph {et~al.}}]{Keesling2018}%
  \BibitemOpen
  \bibfield  {author} {\bibinfo {author} {\bibfnamefont {A.}~\bibnamefont
  {Keesling}}, \bibinfo {author} {\bibfnamefont {A.}~\bibnamefont {Omran}},
  \bibinfo {author} {\bibfnamefont {H.}~\bibnamefont {Levine}}, \bibinfo
  {author} {\bibfnamefont {H.}~\bibnamefont {Bernien}}, \bibinfo {author}
  {\bibfnamefont {H.}~\bibnamefont {Pichler}}, \bibinfo {author} {\bibfnamefont
  {S.}~\bibnamefont {Choi}}, \bibinfo {author} {\bibfnamefont {R.}~\bibnamefont
  {Samajdar}}, \bibinfo {author} {\bibfnamefont {S.}~\bibnamefont {Schwartz}},
  \bibinfo {author} {\bibfnamefont {P.}~\bibnamefont {Silvi}}, \bibinfo
  {author} {\bibfnamefont {S.}~\bibnamefont {Sachdev}},  \emph {et~al.},\
  }\href@noop {} {\bibfield  {journal} {\bibinfo  {journal} {Nature}\ }\textbf
  {\bibinfo {volume} {568}},\ \bibinfo {pages} {207} (\bibinfo {year}
  {2019})}\BibitemShut {NoStop}%
\bibitem [{\citenamefont {Blumoff}\ \emph {et~al.}(2016)\citenamefont
  {Blumoff}, \citenamefont {Chou}, \citenamefont {Shen}, \citenamefont
  {Reagor}, \citenamefont {Axline}, \citenamefont {Brierley}, \citenamefont
  {Silveri}, \citenamefont {Wang}, \citenamefont {Vlastakis}, \citenamefont
  {Nigg}, \citenamefont {Frunzio}, \citenamefont {Devoret}, \citenamefont
  {Jiang}, \citenamefont {Girvin},\ and\ \citenamefont
  {Schoelkopf}}]{Blumoff2016}%
  \BibitemOpen
  \bibfield  {author} {\bibinfo {author} {\bibfnamefont {J.~Z.}\ \bibnamefont
  {Blumoff}}, \bibinfo {author} {\bibfnamefont {K.}~\bibnamefont {Chou}},
  \bibinfo {author} {\bibfnamefont {C.}~\bibnamefont {Shen}}, \bibinfo {author}
  {\bibfnamefont {M.}~\bibnamefont {Reagor}}, \bibinfo {author} {\bibfnamefont
  {C.}~\bibnamefont {Axline}}, \bibinfo {author} {\bibfnamefont {R.~T.}\
  \bibnamefont {Brierley}}, \bibinfo {author} {\bibfnamefont {M.~P.}\
  \bibnamefont {Silveri}}, \bibinfo {author} {\bibfnamefont {C.}~\bibnamefont
  {Wang}}, \bibinfo {author} {\bibfnamefont {B.}~\bibnamefont {Vlastakis}},
  \bibinfo {author} {\bibfnamefont {S.~E.}\ \bibnamefont {Nigg}}, \bibinfo
  {author} {\bibfnamefont {L.}~\bibnamefont {Frunzio}}, \bibinfo {author}
  {\bibfnamefont {M.~H.}\ \bibnamefont {Devoret}}, \bibinfo {author}
  {\bibfnamefont {L.}~\bibnamefont {Jiang}}, \bibinfo {author} {\bibfnamefont
  {S.~M.}\ \bibnamefont {Girvin}}, \ and\ \bibinfo {author} {\bibfnamefont
  {R.~J.}\ \bibnamefont {Schoelkopf}},\ }\href {\doibase
  10.1103/PhysRevX.6.031041} {\bibfield  {journal} {\bibinfo  {journal} {Phys.
  Rev. X}\ }\textbf {\bibinfo {volume} {6}},\ \bibinfo {pages} {031041}
  (\bibinfo {year} {2016})}\BibitemShut {NoStop}%
\bibitem [{\citenamefont {Barends}\ \emph {et~al.}(2016)\citenamefont
  {Barends}, \citenamefont {Shabani}, \citenamefont {Lamata}, \citenamefont
  {Kelly}, \citenamefont {Mezzacapo}, \citenamefont {Las~Heras}, \citenamefont
  {Babbush}, \citenamefont {Fowler}, \citenamefont {Campbell}, \citenamefont
  {Chen} \emph {et~al.}}]{Barends2016}%
  \BibitemOpen
  \bibfield  {author} {\bibinfo {author} {\bibfnamefont {R.}~\bibnamefont
  {Barends}}, \bibinfo {author} {\bibfnamefont {A.}~\bibnamefont {Shabani}},
  \bibinfo {author} {\bibfnamefont {L.}~\bibnamefont {Lamata}}, \bibinfo
  {author} {\bibfnamefont {J.}~\bibnamefont {Kelly}}, \bibinfo {author}
  {\bibfnamefont {A.}~\bibnamefont {Mezzacapo}}, \bibinfo {author}
  {\bibfnamefont {U.}~\bibnamefont {Las~Heras}}, \bibinfo {author}
  {\bibfnamefont {R.}~\bibnamefont {Babbush}}, \bibinfo {author} {\bibfnamefont
  {A.~G.}\ \bibnamefont {Fowler}}, \bibinfo {author} {\bibfnamefont
  {B.}~\bibnamefont {Campbell}}, \bibinfo {author} {\bibfnamefont
  {Y.}~\bibnamefont {Chen}},  \emph {et~al.},\ }\href
  {https://www.nature.com/articles/nature17658} {\bibfield  {journal} {\bibinfo
   {journal} {Nature}\ }\textbf {\bibinfo {volume} {534}},\ \bibinfo {pages}
  {222} (\bibinfo {year} {2016})}\BibitemShut {NoStop}%
\bibitem [{\citenamefont {Otterbach}\ \emph {et~al.}()\citenamefont
  {Otterbach}, \citenamefont {Manenti}, \citenamefont {Alidoust}, \citenamefont
  {Bestwick}, \citenamefont {Block}, \citenamefont {Bloom}, \citenamefont
  {Caldwell}, \citenamefont {Didier}, \citenamefont {Fried}, \citenamefont
  {Hong} \emph {et~al.}}]{Otterbach2017}%
  \BibitemOpen
  \bibfield  {author} {\bibinfo {author} {\bibfnamefont {J.}~\bibnamefont
  {Otterbach}}, \bibinfo {author} {\bibfnamefont {R.}~\bibnamefont {Manenti}},
  \bibinfo {author} {\bibfnamefont {N.}~\bibnamefont {Alidoust}}, \bibinfo
  {author} {\bibfnamefont {A.}~\bibnamefont {Bestwick}}, \bibinfo {author}
  {\bibfnamefont {M.}~\bibnamefont {Block}}, \bibinfo {author} {\bibfnamefont
  {B.}~\bibnamefont {Bloom}}, \bibinfo {author} {\bibfnamefont
  {S.}~\bibnamefont {Caldwell}}, \bibinfo {author} {\bibfnamefont
  {N.}~\bibnamefont {Didier}}, \bibinfo {author} {\bibfnamefont {E.~S.}\
  \bibnamefont {Fried}}, \bibinfo {author} {\bibfnamefont {S.}~\bibnamefont
  {Hong}},  \emph {et~al.},\ }\href {https://arxiv.org/abs/1712.05771}
  {\bibinfo  {journal} {arXiv:1712.05771}\ }\BibitemShut {NoStop}%
\bibitem [{\citenamefont {Gong}\ \emph {et~al.}(2019)\citenamefont {Gong},
  \citenamefont {Chen}, \citenamefont {Zheng}, \citenamefont {Wang},
  \citenamefont {Zha}, \citenamefont {Deng}, \citenamefont {Yan}, \citenamefont
  {Rong}, \citenamefont {Wu}, \citenamefont {Li} \emph {et~al.}}]{Gong2018}%
  \BibitemOpen
\bibfield  {journal} {  }\bibfield  {author} {\bibinfo {author} {\bibfnamefont
  {M.}~\bibnamefont {Gong}}, \bibinfo {author} {\bibfnamefont {M.-C.}\
  \bibnamefont {Chen}}, \bibinfo {author} {\bibfnamefont {Y.}~\bibnamefont
  {Zheng}}, \bibinfo {author} {\bibfnamefont {S.}~\bibnamefont {Wang}},
  \bibinfo {author} {\bibfnamefont {C.}~\bibnamefont {Zha}}, \bibinfo {author}
  {\bibfnamefont {H.}~\bibnamefont {Deng}}, \bibinfo {author} {\bibfnamefont
  {Z.}~\bibnamefont {Yan}}, \bibinfo {author} {\bibfnamefont {H.}~\bibnamefont
  {Rong}}, \bibinfo {author} {\bibfnamefont {Y.}~\bibnamefont {Wu}}, \bibinfo
  {author} {\bibfnamefont {S.}~\bibnamefont {Li}},  \emph {et~al.},\ }\href
  {https://journals.aps.org/prl/abstract/10.1103/PhysRevLett.122.110501}
  {\bibfield  {journal} {\bibinfo  {journal} {Phys. Rev. Lett.}\ }\textbf
  {\bibinfo {volume} {122}},\ \bibinfo {pages} {110501} (\bibinfo {year}
  {2019})}\BibitemShut {NoStop}%
\bibitem [{\citenamefont {Ketterer}\ \emph {et~al.}(2019)\citenamefont
  {Ketterer}, \citenamefont {Wyderka},\ and\ \citenamefont
  {G\"uhne}}]{Ketterer2018}%
  \BibitemOpen
  \bibfield  {author} {\bibinfo {author} {\bibfnamefont {A.}~\bibnamefont
  {Ketterer}}, \bibinfo {author} {\bibfnamefont {N.}~\bibnamefont {Wyderka}}, \
  and\ \bibinfo {author} {\bibfnamefont {O.}~\bibnamefont {G\"uhne}},\ }\href
  {\doibase 10.1103/PhysRevLett.122.120505} {\bibfield  {journal} {\bibinfo
  {journal} {Phys. Rev. Lett.}\ }\textbf {\bibinfo {volume} {122}},\ \bibinfo
  {pages} {120505} (\bibinfo {year} {2019})}\BibitemShut {NoStop}%
\bibitem [{\citenamefont {van Enk}\ and\ \citenamefont
  {Beenakker}(2012)}]{vanEnk:2012}%
  \BibitemOpen
  \bibfield  {author} {\bibinfo {author} {\bibfnamefont {S.~J.}\ \bibnamefont
  {van Enk}}\ and\ \bibinfo {author} {\bibfnamefont {C.~W.~J.}\ \bibnamefont
  {Beenakker}},\ }\href {\doibase 10.1103/PhysRevLett.108.110503} {\bibfield
  {journal} {\bibinfo  {journal} {Phys. Rev. Lett.}\ }\textbf {\bibinfo
  {volume} {108}},\ \bibinfo {pages} {110503} (\bibinfo {year}
  {2012})}\BibitemShut {NoStop}%
\bibitem [{\citenamefont {Nakata}\ \emph {et~al.}(2017)\citenamefont {Nakata},
  \citenamefont {Hirche}, \citenamefont {Koashi},\ and\ \citenamefont
  {Winter}}]{Nakata:2017}%
  \BibitemOpen
  \bibfield  {author} {\bibinfo {author} {\bibfnamefont {Y.}~\bibnamefont
  {Nakata}}, \bibinfo {author} {\bibfnamefont {C.}~\bibnamefont {Hirche}},
  \bibinfo {author} {\bibfnamefont {M.}~\bibnamefont {Koashi}}, \ and\ \bibinfo
  {author} {\bibfnamefont {A.}~\bibnamefont {Winter}},\ }\href {\doibase
  10.1103/PhysRevX.7.021006} {\bibfield  {journal} {\bibinfo  {journal} {Phys.
  Rev. X}\ }\textbf {\bibinfo {volume} {7}},\ \bibinfo {pages} {021006}
  (\bibinfo {year} {2017})}\BibitemShut {NoStop}%
\bibitem [{\citenamefont {Vermersch}\ \emph {et~al.}(2018)\citenamefont
  {Vermersch}, \citenamefont {Elben}, \citenamefont {Dalmonte}, \citenamefont
  {Cirac},\ and\ \citenamefont {Zoller}}]{Vermersch:2018}%
  \BibitemOpen
  \bibfield  {author} {\bibinfo {author} {\bibfnamefont {B.}~\bibnamefont
  {Vermersch}}, \bibinfo {author} {\bibfnamefont {A.}~\bibnamefont {Elben}},
  \bibinfo {author} {\bibfnamefont {M.}~\bibnamefont {Dalmonte}}, \bibinfo
  {author} {\bibfnamefont {J.~I.}\ \bibnamefont {Cirac}}, \ and\ \bibinfo
  {author} {\bibfnamefont {P.}~\bibnamefont {Zoller}},\ }\href {\doibase
  10.1103/PhysRevA.97.023604} {\bibfield  {journal} {\bibinfo  {journal} {Phys.
  Rev. A}\ }\textbf {\bibinfo {volume} {97}},\ \bibinfo {pages} {023604}
  (\bibinfo {year} {2018})}\BibitemShut {NoStop}%
\bibitem [{\citenamefont {Goussev}\ \emph {et~al.}()\citenamefont {Goussev},
  \citenamefont {Jalabert}, \citenamefont {Pastawski},\ and\ \citenamefont
  {Wisniacki}}]{Goussev2012}%
  \BibitemOpen
  \bibfield  {author} {\bibinfo {author} {\bibfnamefont {A.}~\bibnamefont
  {Goussev}}, \bibinfo {author} {\bibfnamefont {R.~A.}\ \bibnamefont
  {Jalabert}}, \bibinfo {author} {\bibfnamefont {H.~M.}\ \bibnamefont
  {Pastawski}}, \ and\ \bibinfo {author} {\bibfnamefont {D.}~\bibnamefont
  {Wisniacki}},\ }\href@noop {} {\ }\Eprint {http://arxiv.org/abs/1206.6348}
  {arXiv:1206.6348} \BibitemShut {NoStop}%
\bibitem [{\citenamefont {Ohliger}\ \emph {et~al.}(2013)\citenamefont
  {Ohliger}, \citenamefont {Nesme},\ and\ \citenamefont
  {Eisert}}]{Ohliger2013}%
  \BibitemOpen
  \bibfield  {author} {\bibinfo {author} {\bibfnamefont {M.}~\bibnamefont
  {Ohliger}}, \bibinfo {author} {\bibfnamefont {V.}~\bibnamefont {Nesme}}, \
  and\ \bibinfo {author} {\bibfnamefont {J.}~\bibnamefont {Eisert}},\ }\href
  {http://iopscience.iop.org/article/10.1088/1367-2630/15/1/015024/meta}
  {\bibfield  {journal} {\bibinfo  {journal} {New J. Phys.}\ }\textbf {\bibinfo
  {volume} {15}} (\bibinfo {year} {2013})}\BibitemShut {NoStop}%
\bibitem [{Note1()}]{Note1}%
  \BibitemOpen
  \bibinfo {note} {We note that this criterium is sufficient but not
  necessary.}\BibitemShut {Stop}%
\bibitem [{\citenamefont {Horodecki}(2003)}]{Horodecki:2003}%
  \BibitemOpen
  \bibfield  {author} {\bibinfo {author} {\bibfnamefont {P.}~\bibnamefont
  {Horodecki}},\ }\href {\doibase 10.1103/PhysRevA.68.052101} {\bibfield
  {journal} {\bibinfo  {journal} {Phys. Rev. A}\ }\textbf {\bibinfo {volume}
  {68}},\ \bibinfo {pages} {052101} (\bibinfo {year} {2003})}\BibitemShut
  {NoStop}%
\bibitem [{\citenamefont {Linke}\ \emph {et~al.}(2018)\citenamefont {Linke},
  \citenamefont {Johri}, \citenamefont {Figgatt}, \citenamefont {Landsman},
  \citenamefont {Matsuura},\ and\ \citenamefont {Monroe}}]{Linke:2017}%
  \BibitemOpen
  \bibfield  {author} {\bibinfo {author} {\bibfnamefont {N.~M.}\ \bibnamefont
  {Linke}}, \bibinfo {author} {\bibfnamefont {S.}~\bibnamefont {Johri}},
  \bibinfo {author} {\bibfnamefont {C.}~\bibnamefont {Figgatt}}, \bibinfo
  {author} {\bibfnamefont {K.~A.}\ \bibnamefont {Landsman}}, \bibinfo {author}
  {\bibfnamefont {A.~Y.}\ \bibnamefont {Matsuura}}, \ and\ \bibinfo {author}
  {\bibfnamefont {C.}~\bibnamefont {Monroe}},\ }\href {\doibase
  10.1103/PhysRevA.98.052334} {\bibfield  {journal} {\bibinfo  {journal} {Phys.
  Rev. A}\ }\textbf {\bibinfo {volume} {98}},\ \bibinfo {pages} {052334}
  (\bibinfo {year} {2018})}\BibitemShut {NoStop}%
\bibitem [{\citenamefont {Gross}\ \emph {et~al.}(2007)\citenamefont {Gross},
  \citenamefont {Audenaert},\ and\ \citenamefont {Eisert}}]{Gross2007}%
  \BibitemOpen
  \bibfield  {author} {\bibinfo {author} {\bibfnamefont {D.}~\bibnamefont
  {Gross}}, \bibinfo {author} {\bibfnamefont {K.}~\bibnamefont {Audenaert}}, \
  and\ \bibinfo {author} {\bibfnamefont {J.}~\bibnamefont {Eisert}},\ }\href
  {https://aip.scitation.org/doi/10.1063/1.2716992} {\bibfield  {journal}
  {\bibinfo  {journal} {J. Math. Phys.}\ }\textbf {\bibinfo {volume} {48}}
  (\bibinfo {year} {2007})}\BibitemShut {NoStop}%
\bibitem [{\citenamefont {Collins}\ and\ \citenamefont
  {Nechita}(2010)}]{Collins2009}%
  \BibitemOpen
  \bibfield  {author} {\bibinfo {author} {\bibfnamefont {B.}~\bibnamefont
  {Collins}}\ and\ \bibinfo {author} {\bibfnamefont {I.}~\bibnamefont
  {Nechita}},\ }\href@noop {} {\bibfield  {journal} {\bibinfo  {journal}
  {Commun. Math. Phys.}\ }\textbf {\bibinfo {volume} {297}},\ \bibinfo {pages}
  {345} (\bibinfo {year} {2010})}\BibitemShut {NoStop}%
\bibitem [{\citenamefont {Dankert}\ \emph {et~al.}(2009)\citenamefont
  {Dankert}, \citenamefont {Cleve}, \citenamefont {Emerson},\ and\
  \citenamefont {Livine}}]{Dankert2009}%
  \BibitemOpen
  \bibfield  {author} {\bibinfo {author} {\bibfnamefont {C.}~\bibnamefont
  {Dankert}}, \bibinfo {author} {\bibfnamefont {R.}~\bibnamefont {Cleve}},
  \bibinfo {author} {\bibfnamefont {J.}~\bibnamefont {Emerson}}, \ and\
  \bibinfo {author} {\bibfnamefont {E.}~\bibnamefont {Livine}},\ }\href
  {\doibase 10.1103/PhysRevA.80.012304} {\bibfield  {journal} {\bibinfo
  {journal} {Phys. Rev. A}\ }\textbf {\bibinfo {volume} {80}},\ \bibinfo
  {pages} {012304} (\bibinfo {year} {2009})}\BibitemShut {NoStop}%
\bibitem [{\citenamefont {Gamel}(2016)}]{Gamel2016}%
  \BibitemOpen
  \bibfield  {author} {\bibinfo {author} {\bibfnamefont {O.}~\bibnamefont
  {Gamel}},\ }\href {\doibase 10.1103/PhysRevA.93.062320} {\bibfield  {journal}
  {\bibinfo  {journal} {Phys. Rev. A}\ }\textbf {\bibinfo {volume} {93}},\
  \bibinfo {pages} {062320} (\bibinfo {year} {2016})}\BibitemShut {NoStop}%
\bibitem [{\citenamefont {Mezzadri}(2007)}]{Mezzadri:2007}%
  \BibitemOpen
  \bibfield  {author} {\bibinfo {author} {\bibfnamefont {F.}~\bibnamefont
  {Mezzadri}},\ }\href {https://arxiv.org/abs/math-ph/0609050} {\bibfield
  {journal} {\bibinfo  {journal} {Notices of the AMS}\ }\textbf {\bibinfo
  {volume} {54}},\ \bibinfo {pages} {592} (\bibinfo {year} {2007})}\BibitemShut
  {NoStop}%
\bibitem [{\citenamefont {Roberts}\ and\ \citenamefont
  {Yoshida}(2017)}]{Roberts2017}%
  \BibitemOpen
  \bibfield  {author} {\bibinfo {author} {\bibfnamefont {D.~A.}\ \bibnamefont
  {Roberts}}\ and\ \bibinfo {author} {\bibfnamefont {B.}~\bibnamefont
  {Yoshida}},\ }\href
  {https://link.springer.com/article/10.1007%2FJHEP04%282017%29121} {\bibfield
  {journal} {\bibinfo  {journal} {J. High Energy Phys.}\ }\textbf {\bibinfo
  {volume} {2017}} (\bibinfo {year} {2017})}\BibitemShut {NoStop}%
\bibitem [{Note2()}]{Note2}%
  \BibitemOpen
  \bibinfo {note} {Intuitively, Haar random unitaries are matrices with
  elements whose real and imaginary parts are independently distributed
  according to a normal distribution, with additional unitary constraints on
  the entire matrix.}\BibitemShut {Stop}%
\bibitem [{\citenamefont {Haake}(2010)}]{Haake2010}%
  \BibitemOpen
  \bibfield  {author} {\bibinfo {author} {\bibfnamefont {F.}~\bibnamefont
  {Haake}},\ }\href {https://books.google.at/books?id=oo03LoIDYQsC} {\emph
  {\bibinfo {title} {Quantum Signatures of Chaos}}},\ Springer Series in
  Synergetics\ (\bibinfo  {publisher} {Springer Berlin Heidelberg},\ \bibinfo
  {year} {2010})\BibitemShut {NoStop}%
\bibitem [{\citenamefont {Pucha{\l}a}\ and\ \citenamefont
  {Miszczak}(2017)}]{Puchala2017}%
  \BibitemOpen
  \bibfield  {author} {\bibinfo {author} {\bibfnamefont {Z.}~\bibnamefont
  {Pucha{\l}a}}\ and\ \bibinfo {author} {\bibfnamefont {J.~A.}\ \bibnamefont
  {Miszczak}},\ }\href {https://doi.org/10.1515/bpasts-2017-0003} {\bibfield
  {journal} {\bibinfo  {journal} {Bull. Polish Acad. Sci. Tech. Sci.}\ }\textbf
  {\bibinfo {volume} {65}},\ \bibinfo {pages} {21} (\bibinfo {year}
  {2017})}\BibitemShut {NoStop}%
\bibitem [{\citenamefont {Brouwer}\ and\ \citenamefont
  {Beenakker}(1996)}]{Brouwer1996}%
  \BibitemOpen
  \bibfield  {author} {\bibinfo {author} {\bibfnamefont {P.~W.}\ \bibnamefont
  {Brouwer}}\ and\ \bibinfo {author} {\bibfnamefont {C.~W.}\ \bibnamefont
  {Beenakker}},\ }\href {\doibase 10.1063/1.531667} {\bibfield  {journal}
  {\bibinfo  {journal} {J. Math. Phys.}\ }\textbf {\bibinfo {volume} {37}},\
  \bibinfo {pages} {4904} (\bibinfo {year} {1996})}\BibitemShut {NoStop}%
\bibitem [{Note3()}]{Note3}%
  \BibitemOpen
  \bibinfo {note} {Note that for clarity, we omitted here the boxes of basis
  states, and just kept the decorations.}\BibitemShut {Stop}%
\bibitem [{\citenamefont {Roy}\ and\ \citenamefont {Scott}(2009)}]{Roy2009}%
  \BibitemOpen
  \bibfield  {author} {\bibinfo {author} {\bibfnamefont {A.}~\bibnamefont
  {Roy}}\ and\ \bibinfo {author} {\bibfnamefont {A.~J.}\ \bibnamefont
  {Scott}},\ }\href
  {https://link.springer.com/article/10.1007%2Fs10623-009-9290-2} {\bibfield
  {journal} {\bibinfo  {journal} {Des. Codes, Cryptogr.}\ }\textbf {\bibinfo
  {volume} {53}},\ \bibinfo {pages} {13} (\bibinfo {year} {2009})}\BibitemShut
  {NoStop}%
\bibitem [{Note4()}]{Note4}%
  \BibitemOpen
  \bibinfo {note} {Loosely speaking, up to the $k$th moment, $k$-designs are as
  random as Haar random unitaries.}\BibitemShut {Stop}%
\bibitem [{\citenamefont {Vermersch}\ \emph {et~al.}()\citenamefont
  {Vermersch}, \citenamefont {Elben}, \citenamefont {Sieberer}, \citenamefont
  {Yao},\ and\ \citenamefont {Zoller}}]{Vermersch2018a}%
  \BibitemOpen
  \bibfield  {author} {\bibinfo {author} {\bibfnamefont {B.}~\bibnamefont
  {Vermersch}}, \bibinfo {author} {\bibfnamefont {A.}~\bibnamefont {Elben}},
  \bibinfo {author} {\bibfnamefont {L.~M.}\ \bibnamefont {Sieberer}}, \bibinfo
  {author} {\bibfnamefont {N.~Y.}\ \bibnamefont {Yao}}, \ and\ \bibinfo
  {author} {\bibfnamefont {P.}~\bibnamefont {Zoller}},\ }\href@noop {} {\
  }\Eprint {http://arxiv.org/abs/arXiv:1807.09087} {arXiv:1807.09087}
  \BibitemShut {NoStop}%
\bibitem [{\citenamefont {Blinder}(2013)}]{Blinder2013}%
  \BibitemOpen
  \bibfield  {author} {\bibinfo {author} {\bibfnamefont {S.~M.}\ \bibnamefont
  {Blinder}},\ }\href {https://books.google.at/books?id=M7TCNAEACAAJ} {\emph
  {\bibinfo {title} {{Guide to Essential Math: A Review for Physics, Chemistry
  and Engineering Students}}}}\ (\bibinfo  {publisher} {Elsevier},\ \bibinfo
  {year} {2013})\BibitemShut {NoStop}%
\bibitem [{\citenamefont {Nie}\ \emph {et~al.}(2019)\citenamefont {Nie},
  \citenamefont {Zhang}, \citenamefont {Zhao}, \citenamefont {Xin},
  \citenamefont {Lu},\ and\ \citenamefont {Li}}]{Nie2019}%
  \BibitemOpen
  \bibfield  {author} {\bibinfo {author} {\bibfnamefont {X.}~\bibnamefont
  {Nie}}, \bibinfo {author} {\bibfnamefont {Z.}~\bibnamefont {Zhang}}, \bibinfo
  {author} {\bibfnamefont {X.}~\bibnamefont {Zhao}}, \bibinfo {author}
  {\bibfnamefont {T.}~\bibnamefont {Xin}}, \bibinfo {author} {\bibfnamefont
  {D.}~\bibnamefont {Lu}}, \ and\ \bibinfo {author} {\bibfnamefont
  {J.}~\bibnamefont {Li}},\ }\href {https://arxiv.org/abs/1903.12237}
  {\bibfield  {journal} {\bibinfo  {journal} {arXiv:1903.12237}\ } (\bibinfo
  {year} {2019})}\BibitemShut {NoStop}%
\bibitem [{\citenamefont {Chen}\ \emph {et~al.}(2012)\citenamefont {Chen},
  \citenamefont {Gu}, \citenamefont {Liu},\ and\ \citenamefont
  {Wen}}]{Wen2012}%
  \BibitemOpen
  \bibfield  {author} {\bibinfo {author} {\bibfnamefont {X.}~\bibnamefont
  {Chen}}, \bibinfo {author} {\bibfnamefont {Z.-C.}\ \bibnamefont {Gu}},
  \bibinfo {author} {\bibfnamefont {Z.-X.}\ \bibnamefont {Liu}}, \ and\
  \bibinfo {author} {\bibfnamefont {X.-G.}\ \bibnamefont {Wen}},\ }\href
  {http://science.sciencemag.org/content/338/6114/1604.long} {\bibfield
  {journal} {\bibinfo  {journal} {Science}\ }\textbf {\bibinfo {volume}
  {338}},\ \bibinfo {pages} {1604} (\bibinfo {year} {2012})}\BibitemShut
  {NoStop}%
\bibitem [{\citenamefont {Pollmann}\ and\ \citenamefont
  {Turner}(2012)}]{Pollmann2012}%
  \BibitemOpen
  \bibfield  {author} {\bibinfo {author} {\bibfnamefont {F.}~\bibnamefont
  {Pollmann}}\ and\ \bibinfo {author} {\bibfnamefont {A.~M.}\ \bibnamefont
  {Turner}},\ }\href
  {https://journals.aps.org/prb/abstract/10.1103/PhysRevB.86.125441} {\bibfield
   {journal} {\bibinfo  {journal} {Phys. Rev. B}\ }\textbf {\bibinfo {volume}
  {86}},\ \bibinfo {pages} {125441} (\bibinfo {year} {2012})}\BibitemShut
  {NoStop}%
\bibitem [{\citenamefont {Haegeman}\ \emph {et~al.}(2012)\citenamefont
  {Haegeman}, \citenamefont {P\'erez-Garc\'{\i}a}, \citenamefont {Cirac},\ and\
  \citenamefont {Schuch}}]{Haegeman2012}%
  \BibitemOpen
  \bibfield  {author} {\bibinfo {author} {\bibfnamefont {J.}~\bibnamefont
  {Haegeman}}, \bibinfo {author} {\bibfnamefont {D.}~\bibnamefont
  {P\'erez-Garc\'{\i}a}}, \bibinfo {author} {\bibfnamefont {I.}~\bibnamefont
  {Cirac}}, \ and\ \bibinfo {author} {\bibfnamefont {N.}~\bibnamefont
  {Schuch}},\ }\href
  {https://journals.aps.org/prl/abstract/10.1103/PhysRevLett.109.050402}
  {\bibfield  {journal} {\bibinfo  {journal} {Phys. Rev. Lett.}\ }\textbf
  {\bibinfo {volume} {109}},\ \bibinfo {pages} {050402} (\bibinfo {year}
  {2012})}\BibitemShut {NoStop}%
\bibitem [{\citenamefont {Shapourian}\ \emph {et~al.}(2017)\citenamefont
  {Shapourian}, \citenamefont {Shiozaki},\ and\ \citenamefont
  {Ryu}}]{Shapourian2017}%
  \BibitemOpen
  \bibfield  {author} {\bibinfo {author} {\bibfnamefont {H.}~\bibnamefont
  {Shapourian}}, \bibinfo {author} {\bibfnamefont {K.}~\bibnamefont
  {Shiozaki}}, \ and\ \bibinfo {author} {\bibfnamefont {S.}~\bibnamefont
  {Ryu}},\ }\href {\doibase 10.1103/PhysRevLett.118.216402} {\bibfield
  {journal} {\bibinfo  {journal} {Phys. Rev. Lett.}\ }\textbf {\bibinfo
  {volume} {118}},\ \bibinfo {pages} {216402} (\bibinfo {year}
  {2017})}\BibitemShut {NoStop}%
\bibitem [{\citenamefont {Li}\ and\ \citenamefont {Haldane}(2008)}]{Hui2008}%
  \BibitemOpen
  \bibfield  {author} {\bibinfo {author} {\bibfnamefont {H.}~\bibnamefont
  {Li}}\ and\ \bibinfo {author} {\bibfnamefont {F.~D.~M.}\ \bibnamefont
  {Haldane}},\ }\href
  {https://journals.aps.org/prl/abstract/10.1103/PhysRevLett.101.010504}
  {\bibfield  {journal} {\bibinfo  {journal} {Phys. Rev. Lett.}\ }\textbf
  {\bibinfo {volume} {101}},\ \bibinfo {pages} {010504} (\bibinfo {year}
  {2008})}\BibitemShut {NoStop}%
\bibitem [{\citenamefont {Pichler}\ \emph {et~al.}(2016)\citenamefont
  {Pichler}, \citenamefont {Zhu}, \citenamefont {Seif}, \citenamefont
  {Zoller},\ and\ \citenamefont {Hafezi}}]{Pichler2016}%
  \BibitemOpen
  \bibfield  {author} {\bibinfo {author} {\bibfnamefont {H.}~\bibnamefont
  {Pichler}}, \bibinfo {author} {\bibfnamefont {G.}~\bibnamefont {Zhu}},
  \bibinfo {author} {\bibfnamefont {A.}~\bibnamefont {Seif}}, \bibinfo {author}
  {\bibfnamefont {P.}~\bibnamefont {Zoller}}, \ and\ \bibinfo {author}
  {\bibfnamefont {M.}~\bibnamefont {Hafezi}},\ }\href
  {https://journals.aps.org/prx/abstract/10.1103/PhysRevX.6.041033} {\bibfield
  {journal} {\bibinfo  {journal} {Phys. Rev. X}\ }\textbf {\bibinfo {volume}
  {6}},\ \bibinfo {pages} {041033} (\bibinfo {year} {2016})}\BibitemShut
  {NoStop}%
\bibitem [{\citenamefont {Dalmonte}\ \emph {et~al.}(2018)\citenamefont
  {Dalmonte}, \citenamefont {Vermersch},\ and\ \citenamefont
  {Zoller}}]{Dalmonte2018}%
  \BibitemOpen
  \bibfield  {author} {\bibinfo {author} {\bibfnamefont {M.}~\bibnamefont
  {Dalmonte}}, \bibinfo {author} {\bibfnamefont {B.}~\bibnamefont {Vermersch}},
  \ and\ \bibinfo {author} {\bibfnamefont {P.}~\bibnamefont {Zoller}},\
  }\href@noop {} {\bibfield  {journal} {\bibinfo  {journal} {Nature Physics}\
  }\textbf {\bibinfo {volume} {14}},\ \bibinfo {pages} {827} (\bibinfo {year}
  {2018})}\BibitemShut {NoStop}%
\bibitem [{\citenamefont {Emerson}\ \emph {et~al.}(2007)\citenamefont
  {Emerson}, \citenamefont {Silva}, \citenamefont {Moussa}, \citenamefont
  {Ryan}, \citenamefont {Laforest}, \citenamefont {Baugh}, \citenamefont
  {Cory},\ and\ \citenamefont {Laflamme}}]{Emerson2007}%
  \BibitemOpen
  \bibfield  {author} {\bibinfo {author} {\bibfnamefont {J.}~\bibnamefont
  {Emerson}}, \bibinfo {author} {\bibfnamefont {M.}~\bibnamefont {Silva}},
  \bibinfo {author} {\bibfnamefont {O.}~\bibnamefont {Moussa}}, \bibinfo
  {author} {\bibfnamefont {C.}~\bibnamefont {Ryan}}, \bibinfo {author}
  {\bibfnamefont {M.}~\bibnamefont {Laforest}}, \bibinfo {author}
  {\bibfnamefont {J.}~\bibnamefont {Baugh}}, \bibinfo {author} {\bibfnamefont
  {D.~G.}\ \bibnamefont {Cory}}, \ and\ \bibinfo {author} {\bibfnamefont
  {R.}~\bibnamefont {Laflamme}},\ }\href {\doibase 10.1126/science.1145699}
  {\bibfield  {journal} {\bibinfo  {journal} {Science}\ }\textbf {\bibinfo
  {volume} {317}},\ \bibinfo {pages} {1893} (\bibinfo {year}
  {2007})}\BibitemShut {NoStop}%
\bibitem [{\citenamefont {Knill}\ \emph {et~al.}(2008)\citenamefont {Knill},
  \citenamefont {Leibfried}, \citenamefont {Reichle}, \citenamefont {Britton},
  \citenamefont {Blakestad}, \citenamefont {Jost}, \citenamefont {Langer},
  \citenamefont {Ozeri}, \citenamefont {Seidelin},\ and\ \citenamefont
  {Wineland}}]{Knill2008}%
  \BibitemOpen
  \bibfield  {author} {\bibinfo {author} {\bibfnamefont {E.}~\bibnamefont
  {Knill}}, \bibinfo {author} {\bibfnamefont {D.}~\bibnamefont {Leibfried}},
  \bibinfo {author} {\bibfnamefont {R.}~\bibnamefont {Reichle}}, \bibinfo
  {author} {\bibfnamefont {J.}~\bibnamefont {Britton}}, \bibinfo {author}
  {\bibfnamefont {R.~B.}\ \bibnamefont {Blakestad}}, \bibinfo {author}
  {\bibfnamefont {J.~D.}\ \bibnamefont {Jost}}, \bibinfo {author}
  {\bibfnamefont {C.}~\bibnamefont {Langer}}, \bibinfo {author} {\bibfnamefont
  {R.}~\bibnamefont {Ozeri}}, \bibinfo {author} {\bibfnamefont
  {S.}~\bibnamefont {Seidelin}}, \ and\ \bibinfo {author} {\bibfnamefont
  {D.~J.}\ \bibnamefont {Wineland}},\ }\href {\doibase
  10.1103/PhysRevA.77.012307} {\bibfield  {journal} {\bibinfo  {journal} {Phys.
  Rev. A}\ }\textbf {\bibinfo {volume} {77}},\ \bibinfo {pages} {012307}
  (\bibinfo {year} {2008})}\BibitemShut {NoStop}%
\bibitem [{\citenamefont {Wallman}\ \emph {et~al.}(2015)\citenamefont
  {Wallman}, \citenamefont {Granade}, \citenamefont {Harper},\ and\
  \citenamefont {Flammia}}]{Wallman2015}%
  \BibitemOpen
  \bibfield  {author} {\bibinfo {author} {\bibfnamefont {J.}~\bibnamefont
  {Wallman}}, \bibinfo {author} {\bibfnamefont {C.}~\bibnamefont {Granade}},
  \bibinfo {author} {\bibfnamefont {R.}~\bibnamefont {Harper}}, \ and\ \bibinfo
  {author} {\bibfnamefont {S.~T.}\ \bibnamefont {Flammia}},\ }\href@noop {}
  {\bibfield  {journal} {\bibinfo  {journal} {New Journal of Physics}\ }\textbf
  {\bibinfo {volume} {17}},\ \bibinfo {pages} {113020} (\bibinfo {year}
  {2015})}\BibitemShut {NoStop}%
\bibitem [{\citenamefont {Johansson}\ \emph {et~al.}(2013)\citenamefont
  {Johansson}, \citenamefont {Nation},\ and\ \citenamefont
  {Nori}}]{Johansson20131234}%
  \BibitemOpen
  \bibfield  {author} {\bibinfo {author} {\bibfnamefont {J.~R.}\ \bibnamefont
  {Johansson}}, \bibinfo {author} {\bibfnamefont {P.~D.}\ \bibnamefont
  {Nation}}, \ and\ \bibinfo {author} {\bibfnamefont {F.}~\bibnamefont
  {Nori}},\ }\href {\doibase 10.1016/j.cpc.2012.11.019} {\bibfield  {journal}
  {\bibinfo  {journal} {Comput. Phys. Commun.}\ }\textbf {\bibinfo {volume}
  {184}},\ \bibinfo {pages} {1234} (\bibinfo {year} {2013})}\BibitemShut
  {NoStop}%
\bibitem [{\citenamefont {Bona}(2016)}]{Bona2016}%
  \BibitemOpen
  \bibfield  {author} {\bibinfo {author} {\bibfnamefont {M.}~\bibnamefont
  {Bona}},\ }\href {https://books.google.at/books?id=Y3nRBQAAQBAJ} {\emph
  {\bibinfo {title} {{Combinatorics of Permutations, Second Edition}}}},\
  Discrete Mathematics and Its Applications\ (\bibinfo  {publisher} {CRC
  Press},\ \bibinfo {year} {2016})\BibitemShut {NoStop}%
\end{thebibliography}
